\documentclass[draftcls,onecolumn,11pt]{IEEEtran}
\usepackage{enumerate,cite,amsmath,amsfonts,amssymb,subfigure,epsfig,times,graphicx,psfrag}
\usepackage{float}
\usepackage{verbatim,color}
\psfull
\newcommand{\mbf}{\boldmath}
\newcommand{\smbf}{\footnotesize \boldmath}

\def\sbav{{\mbox {\smbf $a$}}}

\def\b0{{\mbox {\mbf $0$}}}
\def\bA{{\mbox {\mbf $A$}}}
\def\bav{{\mbox {\mbf $a$}}}

\def\br{{\mbox {\mbf $r$}}}

\def\bv{{\mbox {\mbf $v$}}}

\def\bI{{\mbox {\mbf $I$}}}

\def\bb0{{\mathbf{0}}}
\def\bzero{{\mathbf{0}}}
\def\bone{{\mathbf{1}}}

\def\blambda{\boldsymbol{\lambda}}

\def\bphi{\boldsymbol{\phi}}
\def\bpsi{\boldsymbol{\psi}}
\def\bchi{\boldsymbol{\chi}}

\def\bxi{\boldsymbol{\xi}}

\def\b_beta{\mbox{\boldmath $\beta$}}
\def\hb_beta{\mbox{\boldmath $\hat \beta$}}

\newtheorem{Lemma}{Lemma}

\newtheorem{Remark}{Remark}
\newtheorem{Propo}{\it {Proposition}}
\newtheorem{Property}{Property}
\begin{document}
\vfill
\title{A Simple Sequential Spectrum Sensing Scheme for Cognitive Radio}
\author{Yan Xin$^{\dag}$ and Honghai Zhang$^{\dag}$
\thanks{$^{\dag}$The authors are with NEC Laboratories America, Inc., 4 Independence Way,
Princeton, NJ 08540, USA, tel. no.: 1-609-951-4802, fax no.:
1-609-951-2482, e-mail: \{yanxin, honghai\}@nec-labs.com.}}
\markboth{IEEE TRANSACTIONS ON SIGNAL PROCESSING (SUBMITTED)}{}
\renewcommand{\thepage}{} \maketitle
\begin{center}
{\vspace{2cm}{\bf EDICS}}:\\
SPC-DETC: Detection, estimation, and demodulation\\
SSP-DETC: Detection
\end{center}

\pagenumbering{arabic}
\newpage
\begin{abstract}
Cognitive radio that supports a secondary and opportunistic access
to licensed spectrum shows great potential to dramatically improve
spectrum utilization. Spectrum sensing performed by secondary users
to detect unoccupied spectrum bands, is a key enabling technique for
cognitive radio. This paper proposes a truncated sequential spectrum
sensing scheme, namely the sequential shifted chi-square test
(SSCT). The SSCT has a simple test statistic and does not rely on
any deterministic knowledge about primary signals. As figures of
merit, the {\it exact} false-alarm probability is derived, and the
miss-detection probability as well as the average sample number
(ASN) are evaluated by using a numerical integration algorithm.
Corroborating numerical examples show that, in comparison with
fixed-sample size detection schemes such as energy detection, the
SSCT delivers considerable reduction on the ASN while maintaining a
comparable detection performance.
\end{abstract}
\begin{center}
{\small \bf Key Words}:\\[0.2cm]Cognitive radio, energy detection, hypothesis testing, spectrum
sensing, sequential detection.
\end{center}

\section{Introduction}\label{section:intro}

Most radio frequency spectrum is allocated primarily based on fixed
spectrum allocation strategies that grant licensed users to
exclusively use specific frequency bands to avoid interference.
Recent reports \cite{FCC2002:reportcognitive} released by the
Federal Communications Commission show that, a large amount of
allocated spectrum particularly television bands, is substantially
under-utilized most of the time whereas a small portion of spectrum
bands such as cellular bands, experience increasingly congestion and
scarcity due to rapid deployment of various wireless services.
Cognitive radio, which enables secondary (unlicensed) users to
access licensed spectrum bands not being currently occupied, can
fundamentally alter this unbalanced spectrum usage and therefore can
dramatically improve spectrum utilization. Since licensed (primary)
users are prior to unlicensed (secondary) users in utilizing
spectrum, the secondary and opportunistic access to licensed
spectrum bands is only allowed to have negligible probability of
deteriorating the quality of service (QoS) of primary users (PUs).
Spectrum sensing performed by secondary users (SUs) to detect the
unoccupied frequency bands, is the key enabling technique to meet
this requirement, thereby receiving considerable amount of research
interest recently
\cite{Liang:tradeoff08,Cui07:jssp,seunggg:sqofdm,laifanpoor2008}.

Albeit in essence a conventional signal detection problem, the
design of a spectrum sensing scheme needs to cope with several
critical challenges that stem from special attributes of cognitive
radio networks. First, it is often difficult for SUs in a cognitive
radio network to acquire complete or even partial knowledge about
primary signals. Secondly, SUs need to be able to quickly detect
primary signals at a fairly low detection signal-to-noise ratio
(SNR) level with low detection error probabilities. To this end,
several spectrum sensing schemes, such as matched-filter detection
\cite{conf_icc_chengaodaut}, energy detection
\cite{Liang:tradeoff08,article_energydetectionoriginalpaper}, and
cyclostationary detection \cite{conf_crowncom_lunden}, have been
proposed and investigated. Among these sensing schemes, energy
detection is particularly appealing as it does not rely on any
deterministic knowledge of the primary signals and has low
implementation complexity. However, when the detection SNR is low,
energy detection entails a large amount of sensing time to ensure
high detection accuracy, e.g., the sensing time is inversely
proportional to the square of $\text{SNR}$ \cite{journal:sahai08j}.
To overcome this shortcoming, several sensing schemes based on the
sequential probability ratio test (SPRT) have been proposed under
various cognitive radio settings
\cite{icecs07:Kundargiewfik,chen06:technicalreport,seunggg:sqofdm}.

The SPRT has been widely used in many scientific and engineering
fields since it was introduced by Wald
\cite{IEEEexample:article_sequentialdetection} in 1940s . Perhaps,
the most remarkable character of the SPRT is that, for given
detection error probabilities, the SPRT requires the smallest
average sample number (ASN) for testing simple hypotheses
\cite{WaldWolfowitz:optimumcharacter}. In comparison to
fixed-sample-size sensing schemes, the sensing scheme based on the
SPRT requires much reduced sensing time on the average while
maintaining the same detection performance
\cite{icecs07:Kundargiewfik}. Nonetheless, the existing SPRT based
sensing scheme \cite{icecs07:Kundargiewfik} suffers from several
potential drawbacks. First, when primary signals are taken from a
finite alphabet, the test statistic involves a special function,
which incurs high implementation complexity
\cite{icecs07:Kundargiewfik,seunggg:sqofdm}. Second, evaluating the
probability ratio requires deterministic knowledge or statistical
distribution of certain parameters of the primary signals. Acquiring
such deterministic information or statistical distribution is
practically difficult in general. Thirdly, the existing SPRT based
sensing scheme adopts the Wald's choice on the thresholds
\cite{IEEEexample:article_sequentialdetection}. However, the Wald's
choice, which works well for the non-truncated SPRT, increases error
probabilities when applied to the truncated SPRT.

In this paper, we propose a truncated sequential spectrum sensing
scheme, namely the sequential shifted chi-square test (SSCT). The
SSCT possesses several attractive features: 1) Like energy
detection, the SSCT only requires the knowledge on noise power and
does not rely on any deterministic knowledge about primary signals;
2) compared to fixed-sample-size detection such as energy detection,
the SSCT is capable of delivering considerable reduction on the
average sensing time while maintaining a comparable detection
performance; 3) in comparison with the SPRT based sensing scheme
\cite{icecs07:Kundargiewfik}, the SSCT has a much simpler test
statistic and thus has lower implementation complexity;  4) and the
SSCT offers desirable flexibility to strike a trade-off between
detection performance and sensing time when the operating SNR is
higher than the minimum detection SNR. To evaluate the detection
performance of the SSCT, we derive the exact false-alarm
probability, and employ a numerical integration algorithm from
\cite{pollockgolhar:technicalreport} to compute the miss-detection
probability and the ASN in a recursive manner. Notably, the problem
of evaluating the false-alarm probability of the SSCT resembles the
exact operating characteristic (OC) evaluation problem associated
with truncated sequential life tests involving the exponential
distribution
\cite{IEEEexample:article_exacttruncatedsequential}\cite{Aroian68:directmethod}.
The latter problem has been solved by Woodall and Kurkjian
\cite{IEEEexample:article_exacttruncatedsequential}. Despite the
similarity of these two problems, the Woodall-Kurkjian approach
cannot be applied to directly evaluate the false-alarm probability
of the SSCT. In addition, the Woodall-Kurkjian approach is not
applicable to evaluate the ASN for truncated sequential life tests
involving exponential distribution \cite{Aroian68:directmethod}. As
a byproduct, our approach to evaluating the false-alarm probability
of the SSCT, can be readily modified to evaluate the ASN for
truncated sequential life tests in the exponential case.

The remainder of this paper is organized as follows. Section
\ref{section:model} presents the problem formulation and provides
necessary preliminaries on energy detection and a SPRT based sensing
scheme. Section \ref{section:detectionscheme} introduces the SSCT
and its equivalent test procedure. Section
\ref{section:detectionperformance} deals with the evaluation of the
error probabilities of the SSCT. In particular, this section
provides an exact result for the false-alarm probability, and a
numerical integration algorithm to recursively compute the
miss-detection probability. Section \ref{section:averagesensingtime}
presents an evaluation result on the ASN of the SSCT, while Section
\ref{section:simulation} provides several numerical examples.
Finally, Section \ref{section:conclusion} concludes the paper.

The following notation is used in this paper. Boldface upper and
lower case letters are used to denote matrices and vectors,
respectively; $\bI_k$ denotes a $k\times k$ identity matrix;
$E[\cdot]$ denotes the expectation operator. $(\cdot)^T$ denotes the
transpose operation; $\bone_k$ denotes a $k\times 1$ vector whose
entries are all ones; $\aleph_p^q$ denotes a set of consecutive
integers from $p$ to $q$, i.e., $\aleph_p^q:=\{p,p+1,\ldots,q\}$,
where $p$ is a non-negative integer and $q$ is a positive integer or
infinity; $(\cdot)^c$ denotes a complement of a set;
${\mathbb{I}}_{\{x\ge t\}}$ denotes an indicator function defined as
${\mathbb{I}}_{\{x\ge t\}}=1$ if $x \ge t$ and ${\mathbb{I}}_{\{x\ge
t\}}=0$ if $x<t$, where $x$ is a variable and $t$ is a constant.

\section{Problem Formulation and Preliminaries}\label{section:model}

In this section, we start by presenting a statistical formulation of
the spectrum sensing problem for a single SU cognitive radio system.
We next give a brief overview on two sensing schemes, energy
detection and a SPRT based sensing scheme, which are closely related
to the SSCT.

\subsection{Problem Formulation}\label{subsection:systemmodel}

Consider a narrow-band cognitive radio communication system having a
single SU. The SU shares the same spectrum with a single PU and
needs to detect the presence/absence of the PU. Let $H_0$ and $H_1$
denote the null and alternative hypotheses, respectively.  The
detection of the primary signals can be formulated as a binary
hypothesis testing problem as follows
\begin{align}\label{eq:hypothesistesting}
&H_0: r_i=w_i,~i=1,2,\ldots,\\
&H_1: r_i=h s_i+w_i,~i=1,2,\ldots,
\end{align}
where $r_i$ is the received signal at the SU, $w_i$ is additive
white Gaussian noise, $h$ is the channel gain between the PU and the
SU, and $s_i$ is the transmitted signal of the PU. We further assume
that
\begin{enumerate}
\item [A1)] $w_i$'s are modeled as independent and identically
distributed (i.i.d.) zero mean complex Gaussian random variables
(RVs) with variance $\sigma_w^2/2$ per dimension, i.e., $w_i\sim
\mathcal{CN}(0,\sigma_w^2)$, \item [A2)] the channel gain $h$ is
constant during the sensing period,
\item [A3)] the primary signal samples $s_i$ are i.i.d.,
\item [A4)] $w_i$ and $s_i$ are statistically independent, \item
[A5)] and the perfect knowledge on the noise variance (noise power)
$\sigma_w^2$ is available at the SU.
\end{enumerate}

\subsection{Preliminaries}\label{subsection:preliminary}

\subsubsection{Energy
Detection}\label{subsubsection:energydetection}

In energy detection, the energy of the received signal samples is
first computed and then is compared to a predetermined threshold.
The test procedure of energy detection is given as
\[
T(\br)=\sum_{i=1}^M |r_i|^2 \underset{H_0}{\overset{H_1}\gtreqqless}
\gamma_{\text{ed}}
\]
where $\br:=[r_1,r_2,\ldots,r_M]$, $T(\br)$ denotes the test
statistic, $M$ denotes the fixed sample size, and
$\gamma_{\text{ed}}$ denotes a threshold for energy detection.

Recall that $w_i$ is a zero mean complex Gaussian RV with variance
$\sigma_w^2/2$ per dimension. Under $H_0$, the RV
$2T(\br)/\sigma_w^2$ is a central chi-square RV with $2M$ degrees of
freedom whereas under $H_1$, the RV $2T(\br)/\sigma_w^2$
conditioning on $|s_i|^2~,i=1,\ldots,M$, is a noncentral chi-square
RV with $2M$ degrees of freedom and non-centrality parameter $2|h|^2
\sum_{i=1}^M |s_i|^2/\sigma_w^2$. As $M$ increases, $2|h|^2
\sum_{i=1}^M |s_i|^2/\sigma_w^2$ approaches
$2M|h|^2\sigma_s^2/\sigma_w^2$, where $\sigma_s^2$ denotes the
average symbol energy. Let us define
${\tt{SNR}}_m:=|h|^2\sigma_s^2/\sigma_w^2$ as the minimum detection
SNR. Notice that the minimum detection ${\tt{SNR}}_m$ is a design
parameter referring to the minimum SNR value by which a detector can
achieve the target false-alarm and miss-detection probabilities. In
practice, it is highly likely that ${\tt{SNR}}_m$ is different from
the exact operating SNR, which is typically difficult to acquire in
practice. To distinguish these two different SNRs, we denote by
${\tt{SNR}}_o$ the operating SNR.

It follows directly from the central limit theorem (CLT) that as $M$
approaches infinity, the distribution of $2T(\br)/\sigma_w^2$
converges to a normal distribution given as follows
\cite{Cui07:jssp}
\begin{align}\label{eq:trnormalize}
\frac{2T(\br)}{\sigma_w^2} \sim \begin{cases} \mathcal{N}(2M,4M), & \text{under}~H_0, \\
\mathcal{N}\big(2M(1+{\tt{SNR}}_m), 4M(1+2{\tt{SNR}}_m)\big), &
\text{under}~H_1.
\end{cases}
\end{align}

Typically, the required sample size $M$ is determined by the target
false-alarm and miss-detection probabilities, which we denote by
$\bar \alpha_{\text{ed}}$ and $\bar \beta_{\text{ed}}$ respectively.
Let $M_{\text{ed}}^{\min}$ be the minimum sample number required to
achieve the target $\bar \alpha_{\text{ed}}$ and $\bar
\beta_{\text{ed}}$ at the detection ${\tt{SNR}}_{m}$ level. As shown
in \cite{Liang:tradeoff08}, we have
\begin{equation}\label{eq:minimumsamples}
M_{\text{ed}}^{\min}={\tt{Ceil}}
\Big(({\tt{SNR}}_{m})^{-2}\Big[\mathcal Q^{-1}(\bar
\alpha_{\text{ed}})-\mathcal
Q^{-1}(1-\bar\beta_{\text{ed}})\sqrt{2{\tt{SNR}}_{m}+1}
\Big]^2\Big)=O({\tt{SNR}}_m^{-2})
\end{equation}
where ${\tt{Ceil}}(x)$ denotes the smallest integer not less than
$x$, $\mathcal Q (\cdot)$ is the complementary cumulative
distribution function of the standard normal RV, i.e.,
$\mathcal{Q}(x):=(2\pi)^{-1/2}\int_{x}^{\infty} e^{-t^2/2}dt$, and
$\mathcal Q^{-1}(\cdot)$ denotes its inverse function. It is evident
from \eqref{eq:minimumsamples} that for energy detection, the number
of the required sensing samples is inversely proportional to
${\tt{SNR}}_m^{2}$ when ${\tt{SNR}}_m$ is sufficiently small
\cite{journal:sahai08j}.

As clear from the above description, energy detection has a simple
test statistic and has low implementation complexity
\cite{article_energydetectionoriginalpaper}. In addition, known as a
form of non-coherent detection, energy detection only requires the
knowledge on noise power and does not rely on any deterministic
knowledge about the primary signals $s_i$. However, one major
drawback of energy detection is that, at a low detection SNR level,
it requires a large amount of sensing time to achieve low detection
error probabilities.

\subsubsection{A SPRT Based Sensing Scheme}\label{subsubsection:sprt}

In comparison with a fixed-sample-size detection such as energy
detection, the SPRT is capable of achieving the same detection
performance with a much reduced ASN
\cite{WaldWolfowitz:optimumcharacter}. We next investigate a SPRT
based sensing scheme that relies on the amplitude squares of the
received signal samples \cite{seunggg:sqofdm,icecs07:Kundargiewfik}.
To simplify our description, we now assume that the amplitude
squares of primary signals, $|s_i|^2,~i=1,2,\ldots$, are perfectly
known at the SU. With this assumption, the spectrum sensing problem
formulated in \eqref{eq:hypothesistesting} becomes a simple
hypothesis testing problem, which is the original setting considered
by Wald \cite{IEEEexample:article_sequentialdetection}.

We normalize $|r_i|^2$ as $v_i:=2|r_i|^2/\sigma_w^2$ for the
convenience of derivation. Note that under $H_0$, $v_i$ is an
exponential RV with rate parameter $1/2$ and under $H_1$, $v_i$
conditional on $|s_i|^2$ is a noncentral chi-square RV with two
degrees of freedom and non-centrality parameter $\lambda_i$ that can
be readily obtained as $\lambda_i=2|h|^2|s_i|^2/\sigma_w^2$. Hence,
the probability density function (PDF) of $v_i$ under $H_0$ is
\begin{equation}\label{eq:pdfh0}
p_{H_0}(v_i)=\frac{1}{2}e^{-v_i/2}
\end{equation}
whereas under $H_1$, the PDF of $v_i$ conditional on $\lambda_i$
is
\begin{equation}\label{eq:pdfh1}
p_{H_1}(v_i|\lambda_i)=\frac{1}{2}e^{-(v_i+\lambda_i)/2}I_0(\sqrt{\lambda_i
v_i})
\end{equation}
where $I_0(\cdot)$ is the zeroth-order modified Bessel function of
the first kind. After collecting $N$ samples, we can express the
accumulative log-likelihood ratio as
\begin{align}\label{eq:sequential}
L_N(\bv_N|\blambda_N)&=\log
\frac{p_{\bv|H_1}(\bv_N|\blambda_N)}{p_{\bv|H_0}(\bv_N)}=\sum_{i=1}^N
z_i=-\sum_{i=1}^N \lambda_i/2+\sum_{i=1}^N \log I_0(\sqrt{\lambda_i
v_i})
\end{align}
where $\bv_N:=(v_1,v_2,\ldots,v_N)$, $z_i=\log
\big(p_{H_1}(v_i|\lambda_i)/p_{H_0}(v_i)\big)$,
 and
$\blambda_N:=(\lambda_1,\lambda_2,\ldots,\lambda_N)$. The test
procedure is given as follows: Reject $H_0$, if
$L_N(\bv_N|\blambda_N) \ge b_L$; Accept $H_0$, if
$L_N(\bv_N|\blambda_N) \le a_L$; and continue sensing, if $a_L<
L_N(\bv_N|\blambda_N) < b_L$. In
\cite{IEEEexample:article_sequentialdetection}, Wald specified a
particular choice of the thresholds $a_L$ and $b_L$ for the
non-truncated SPRT as follows
\begin{equation}\label{eq:choice}
a_L=\log\frac{\bar\beta_{\text{sprt}}}{1-\bar\alpha_{\text{sprt}}},~\text{and}
\quad
b_L=\log\frac{1-\bar\beta_{\text{sprt}}}{\bar\alpha_{\text{sprt}}}
\end{equation}
where $\bar\alpha_{\text{sprt}}$ and $\bar\beta_{\text{sprt}}$
denote the target false-alarm and miss-detection probabilities,
respectively. For a non-truncated SPRT, the Wald's choice on $a_L$
and $b_L$ yields true false-alarm and miss-detection probabilities
that are fairly close to the target ones.

Let $z$ be a RV that has the same PDF as $z_i$. It has been
pointed out in \cite{surveyjohnson:sequential} that, in the SPRT,
if hypotheses $H_0$ and $H_1$ are distinct, then
$E_{H_0}(z)<0<E_{H_1}(z)$,  where $E_{H_i}(\cdot)$ denotes the
conditional expectation under $H_i,~i=1,2$.

As evident from \eqref{eq:sequential}, one shortcoming of this
SPRT-based sensing scheme is that the test statistic contains a
modified Bessel function, which may result in high implementation
complexity. When the perfect knowledge of the instantaneous
amplitude squares of the primary signals is not available, the PDF
under $H_1$ is not completely known, i.e., the alternative
hypothesis is composite. Generally speaking, two approaches, the
Bayesian approach or the generalized likelihood ratio test, can be
used to deal with such a case. In the Bayesian approach, {\it a
prior} PDF of the amplitude squares of the primary signals is
required and multiple summations over all possible amplitudes of the
primary signals need to be performed, whereas in the generalized
likelihood ratio test, a maximum likelihood estimation (MLE) of the
amplitude squares of the primary signals is needed \cite{limth08}.
Either of these two approaches, however, leads to a considerable
increase in implementation complexity.

\section{A Simple Sequential Spectrum Sensing Scheme}\label{section:detectionscheme}

We now present a simple sequential spectrum sensing scheme with the
following test statistic
\begin{equation}\label{eq:decisionmetric}
\Lambda_N=\sum_{i=1}^N\left(|r_i|^2-\Delta \right)
\end{equation}
where $\Delta$ is a predetermined constant. Assuming that the
detector needs to make a decision within $M$ samples, we propose the
following test procedure
\begin{align}
\mathtt{Reject}~H_0:~& \text{if}~\Lambda_N \ge b~\text{and}~N \le M-1,~\text{or~if}~\Lambda_{M} \ge \gamma;\label{eq:h1rule1} \\
\mathtt{Accept}~H_0:~& \text{if}~\Lambda_N \le a~\text{and}~N \le M-1,~ \text{or~if} ~\Lambda_{M} < \gamma;~\label{eq:h0rule1} \\
\mathtt{Continue~Sensing:}~& \text{if}~\Lambda_N\in
(a,b)~\text{and}~N \le M-1 \label{eq:continuesample}
\end{align}
where $a$, $b$, and $\gamma$ are three predetermined thresholds with
$a<0$, $b>0$, and $\gamma\in (a,b)$.

\begin{figure}[h]
\centering \psfrag{a}{\footnotesize{$a$}}
\psfrag{b}{\footnotesize{$b$}}
\psfrag{\gamma}{\footnotesize{$\gamma$}}\psfrag{M}{\footnotesize{$M$}}\psfrag{N}{\footnotesize{$N$}}
\psfrag{lambdan}{\footnotesize{$\Lambda_N$}}
\psfrag{Reject}{\footnotesize{$\mathtt{Reject}~H_0$}}
\psfrag{Number}{\footnotesize{$\mathtt{Sample~Index}$}}
\psfrag{Accept}{\footnotesize{$\mathtt{Accept}~H_0$}}
\psfrag{x}{\footnotesize{$\bigstar$}} \psfrag{Continue
Sampling}{\footnotesize{$\mathtt{Continue ~Sampling}$}}
\psfrag{test}{\footnotesize{$\mathtt{Test~Statistic}$}}
\epsfig{file=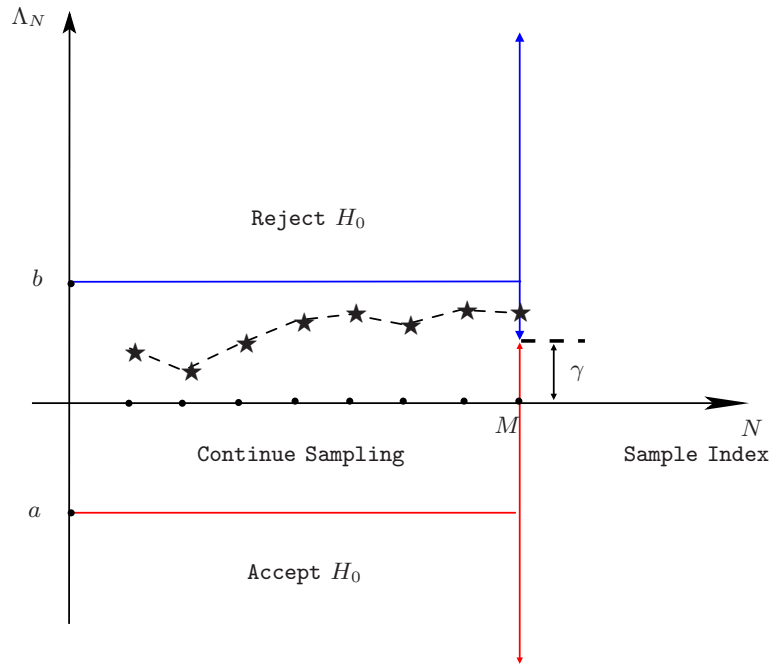,width=0.6\textwidth} \caption{The
test region of the SSCT.} \label{fig:testregionoriginal}
\end{figure}

In statistical term, the test procedure given in
\eqref{eq:h1rule1}--\eqref{eq:continuesample} is nothing but a {\it
truncated} sequential test. As depicted in Fig.
\ref{fig:testregionoriginal}, the stopping boundaries of the test
region consist of a horizonal line $b$ and a horizonal line $a$,
which we simply call the upper- and lower-boundary respectively.
Since each term in the cumulative sum $\Lambda_N$ is a shifted
chi-square RV, we simply term the test procedure
\eqref{eq:h1rule1}-\eqref{eq:continuesample}, {\it the sequential
shifted chi-square test}. It is evident from
\eqref{eq:h1rule1}--\eqref{eq:continuesample} that the test
statistic depends only on the amplitude squares of the received
signal samples and the constant $\Delta$.

Let $r$ be a RV having the same PDF as $|r_i|^2-\Delta$, which is
the $i$th incremental term in the test statistic
\eqref{eq:decisionmetric}. As we show below, we choose $\sigma_w^2 <
\Delta < \sigma_w^2(1+{\tt{SNR}}_m)$ to ensure
$E_{H_0}(r)<0<E_{H_1}(r)$ similar to the SPRT case. In the SSCT, we
have $E_{H_0}(r)=\sigma_w^2-\Delta$ and
$E_{H_1}(r)=\sigma_w^2(1+{\tt{SNR}}_o)-\Delta$. Using
${\tt{SNR}}_m\le {\tt{SNR}}_o$ and $\sigma_w^2 < \Delta <
\sigma_w^2(1+{\tt{SNR}}_m)$, we always have $E_{H_0}(r)<0 \le
{\tt{SNR}}_o-{\tt{SNR}}_m < E_{H_1}(r)$. Note that with this choice,
the constant $\Delta$ depends on the minimum detection SNR instead
of the exact operating SNR.

\begin{figure}[h]
\centering \psfrag{an}{\footnotesize{$a_N$}}
\psfrag{a1}{\footnotesize{$a_1$}}\psfrag{ap}{\footnotesize{${a_P}$}}
\psfrag{aM}{\footnotesize{$a_M$}} \psfrag{bM}{\footnotesize{$b_M$}}
\psfrag{a}{\footnotesize{$\bar a$}} \psfrag{b}{\footnotesize{$\bar
b$}}
\psfrag{b1}{\footnotesize{$b_1$}}\psfrag{b2}{\footnotesize{$b_2$}}
\psfrag{an}{\footnotesize{$a_N$}}\psfrag{bn}{\footnotesize{$b_N$}}
\psfrag{gammaM}{\footnotesize{$\bar{\gamma}_M$}}\psfrag{M}{\footnotesize{$M$}}\psfrag{N}{\footnotesize{$N$}}
\psfrag{lambdaN}{\footnotesize{$\xi_N$}}
\psfrag{xi1}{\footnotesize{$\xi_1$}}\psfrag{xi2}{\footnotesize{$\xi_2$}}
\psfrag{xiN1}{\footnotesize{$\xi_{N-1}$}}\psfrag{xiN}{\footnotesize{$\xi_N$}}
\psfrag{N}{\footnotesize{$N$}} \psfrag{M}{\footnotesize{$M$}}
\psfrag{Reject}{\footnotesize{$\mathtt{Reject}~H_0$}}
\psfrag{Accept}{\footnotesize{$\mathtt{Accept}~H_0$}}
\psfrag{Number}{\footnotesize{$\mathtt{Sample~Index}$}}
\psfrag{gamma}{\footnotesize{$\bar\gamma$}}
\psfrag{x}{\footnotesize{$\bigstar$}}
\psfrag{teststatistic}{\footnotesize{$\mathtt{Test~Statistic}$}}
\psfrag{Continue Sampling}{\footnotesize{$\mathtt{Continue
~Sampling}$}} \epsfig{file=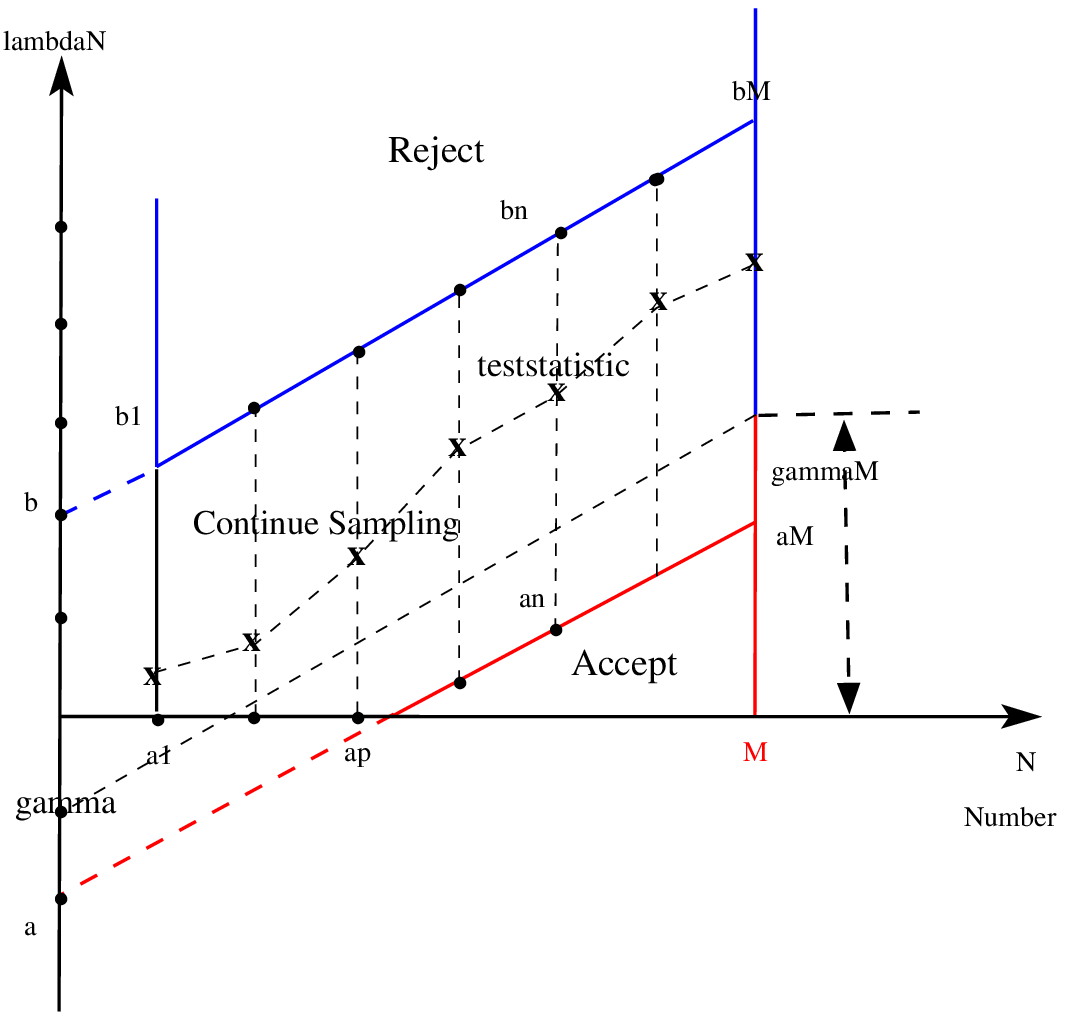,width=0.7\textwidth}
\caption{The test region of the transformed test procedure.}
\label{fig:testregiontransformed}
\end{figure}

We normalize the test statistic $\Lambda_N$ by $\sigma_w^2/2$ and
rewrite \eqref{eq:decisionmetric} as
\begin{equation}\label{eq:lambdabar}
\bar{\Lambda}_N=\sum_{i=1}^N (v_i-2\Delta/\sigma_w^2)
\end{equation}
where $\bar{\Lambda}_N:=2\Lambda_N/\sigma_w^2$ and
$v_i:=2|r_i|^2/\sigma_w^2$. Let $\xi_N$ denote the sum of $v_i$
for $i=1,\ldots,N$, i.e., $\xi_N=\sum_{i=1}^N v_i$ and let
$\bar\Delta$ denote $2\Delta/\sigma_w^2$. With this notation, we
can rewrite $\bar{\Lambda}_N$ as
\begin{equation}\label{eq:transformation}
\bar{\Lambda}_N=\xi_N-N\bar\Delta.
\end{equation}
For notional convenience, we define $\bar{\Lambda}_0$ and $\xi_0$ as
zero. Let $a_i$ and $b_i$ be two parameters defined as follows:
$a_i=0$ for $\aleph_0^P$, $a_i=\bar a+i \bar\Delta $ for $i\in
\aleph_{P+1}^{\infty}$, and $b_i=\bar b+i \bar\Delta$ for $b\in
\aleph_0^{\infty}$, where $\bar a:=2a/\sigma_w^2$, $\bar
b:=2b/\sigma_w^2$, and $P$ denotes the largest integer not greater
than $-a/\Delta$, i.e., $P:={\tt{floor}}(-a/\Delta)$. Using
$\xi_N=\sum_{i=1}^N v_i$, we rewrite the test procedure
\eqref{eq:h1rule1}--\eqref{eq:continuesample} as
\begin{align}
\mathtt{Reject}~H_0:~& \mathtt{if}~\xi_N \ge b_N~\mathtt{and}~N \le
M-1,
 ~\mathtt{or~if}~\xi_{M} \ge \bar{\gamma}_M; \label{eq:h1rule1transform} \\
\mathtt{Accept}~H_0:~& \mathtt{if}~\xi_N \le a_N~\mathtt{and}~N \le
M-1,
~ \mathtt{or~if} ~\xi_{M} < \bar{\gamma}_M;~ \label{eq:h0rule1transform} \\
\mathtt{Continue~Sensing}:~& \mathtt{if}~\xi_N\in
(a_N,b_N)~\mathtt{and}~N \le M-1 \label{eq:continuesampletransform}
\end{align}
where $\bar{\gamma}_M=\bar\gamma+M\bar \Delta$ with
$\bar\gamma:=2\gamma/\sigma_w^2$. The corresponding test region is
depicted in Fig. \ref{fig:testregiontransformed}, where the stopping
boundaries comprise two slant line segments. We adopt
$\alpha_{\text{ssct}}$ and $\beta_{\text{ssct}}$ to denote the
false-alarm and miss-detection probabilities of the SSCT,
respectively.

Since the proposed test procedure in
\eqref{eq:h1rule1}-\eqref{eq:continuesample} is not necessarily a
SPRT, the Wald's choice on thresholds, which yields a non-truncated
SPRT satisfying specified false-alarm and miss-detection
probabilities, is no longer applicable. Alternatively and
conventionally, the thresholds $a$, $b$, $\gamma$, the parameter
$\Delta$, and a truncated size $M$ are selected beforehand, either
purposefully or randomly, and corresponding $\alpha_{\text{ssct}}$
and $\beta_{\text{ssct}}$ are then computed. If the resulted
$\alpha_{\text{ssct}}$ and $\beta_{\text{ssct}}$ do not meet the
requirement, the thresholds and truncated size are subsequently
adjusted. This process continues until a desirable error probability
performance is obtained. In the above process, the key step is to
accurately and efficiently evaluate the false-alarm and
miss-detection probabilities as well as the ASN for prescribed
thresholds $a$, $b$, $\gamma$, the parameter $\Delta$, and a
truncated size $M$, as will be addressed in the following section.

\section{Evaluations of False-Alarm and Miss-Detection Probabilities}\label{section:detectionperformance}

This section presents the exact false-alarm probability, and a
numerical integration algorithm that obtains the miss-detection
probability in a recursive manner
\cite{pollockgolhar:technicalreport}. We start by introducing some
preparatory tools, including three mutually related integrals that
will be frequently used in the evaluation of the false-alarm
probability.

\subsection{Preparatory Tools}\label{subsect:preparatorytools}

The first integral is defined as
\begin{align}\label{eq:giu}
f^{(k)}_{\bchi_k}(\xi)=1,~k=0,~\text{and}~f^{(k)}_{\bchi_k}(\xi)=
\int^{\xi}_{\chi_{k}}d\xi_{k}\int^{\xi_{k}}_{\chi_{k-1}}d\xi_{k-1}\cdots\int^{\xi_2}_{\chi_1}d\xi_1,~k\ge
1
\end{align}
with $\bchi_0=\emptyset$ and
$\bchi_k:=[\chi_1,\ldots,\chi_{k-1},\chi_k]$ with $0\le \chi_1 \le
\ldots \le \chi_k$. Note that superscript $k$ and subscript
$\bchi_k$ are used to indicate that $f^{(k)}_{\bchi_k}(\xi)$ is a
$k$-fold multiple integral with ordered lower limits specified by
$\bchi_k$. Evidently, the integral $f_{\bchi_k}^{(k)}(\xi)$ is a
polynomial in $\xi$ of degree $k$. The following lemma shows that
the exact value of the integral $f^{(k)}_{\bchi_k}(\xi)$ can be
obtained in a recursive manner (see Appendix \ref{proof:lemma1} for
the proof).
\begin{Lemma}\label{Lemma:lemma1}
The integral $f_{\bchi_k}^{(k)}(\xi)$ is given by
\begin{equation}\label{eq:fchikalternative}
f_{\bchi_k}^{(k)}(\xi)=\sum_{i=0}^{k-1}
\frac{f_i^{(k)}(\xi-\chi_{i+1})^{k-i}}{(k-i)!}+f_k^{(k)}
\end{equation}
where the coefficients $f_i^{(k)},~i=0,\ldots,k$, for $k\ge 1$ can
be computed by using the following recurrence relation
\begin{align}\label{eq:recursive1}
f_i^{(k)}=f_i^{(k-1)},~i\in
\aleph_0^{k-1},~f_k^{(k)}=-\sum_{i=0}^{k-1}\frac{f_i^{(k-1)}}{(k-i)!}(\chi_k-\chi_{i+1})^{k-i}
\end{align}
and the initial conditions $f_0^{(0)}=1$. In particular, if
$\chi_1=\chi_2=\ldots=\chi_k=\chi$, $f^{(k)}_{\bchi_k}$ is given
by
\begin{equation}\label{eq:specialcasefk}
f^{(k)}_{\bchi_k}=\frac{1}{k!}(\xi-\chi)^k.
\end{equation}
Furthermore, the integral $f^{(k)}_{\bchi_k}(\xi)$ satisfies the
following properties
\begin{enumerate}
\item {\it Differential Property:}
$df^{(k)}_{\bchi_k}(\xi)/d\xi=f^{(k-1)}_{\bchi_{k-1}}(\xi)$ with
$\bchi_{k-1}=[\chi_1,\ldots,\chi_{k-1}]$ and $k\ge 2$; \item {\it
Scaling Property:}
$f^{(k)}_{t\bchi_k}(t\xi)=t^kf_{\bchi_k}^{(k)}(\xi)$ for $t>0$;
\item {\it Shift Property:}\quad
$f^{(k)}_{\bchi_k-\delta\bone_k}(\xi-\delta)=f^{(k)}_{\bchi_k}(\xi)$.
\end{enumerate}
\end{Lemma}

It is noteworthy to mention that scaling and shift properties are
particularly useful in reducing round-off errors when evaluating
$f^{(k)}_{\bchi_k}(\xi)$. We now introduce the second integral as
\begin{equation}
I^{(0)}:=1,~\text{and}~
I^{(n)}:=\idotsint\limits_{\Omega^{(n)}}d\bxi_n,~n\ge
1~\label{eq:integralI}
\end{equation}
where $\bxi_n:=[\xi_1,\xi_2,\ldots,\xi_n]$ with $0\le \xi_1 \le
\xi_2 \cdots \le \xi_n$, and
$\Omega^{(n)}=\{(\xi_1,\xi_2,\ldots,\xi_{n}): 0\le \xi_1 \le \cdots
\le \xi_{n}; a_i<\xi_i<b_i,i \in \aleph_1^n\}$. In particular, when
$n=1$, $I^{(1)}=\int_{a_1}^{b_1} d\xi_1=b_1-a_1$. Let $c$ and $d$
denote two non-negative real numbers satisfying $0\le c<d$,
$a_{N-1}\le c\le b_N$ and $a_N\le d$. Define the following vector
\begin{align*}
\bpsi_{n,c}^N=\begin{cases}
[\underset{Q}{\underbrace{b_{n+1},\ldots,b_{n+1}}},\underset{N-Q-n}
{\underbrace{a_{Q+n+1},\ldots,a_{N-1},c}}],&n\in
\aleph_0^{N-Q-2}, \\
[\underset{N-n}{\underbrace{b_{n+1},\ldots,b_{n+1},c}}],&n\in \aleph_{N-Q-1}^{s-1},\\
b_{n+1}\bone_{N-n},&n \in \aleph_{s}^{N-2},
\end{cases}
\end{align*}
where $s$ denotes the integer such that $b_s < c \le b_{s+1}$, $Q$
denotes the integer such that $a_Q \le b_1 < a_{Q+1}$, and $N\ge 2$.
Let $\bA_i$ be an $(N-n)\times (N-n-i)$ matrix defined as
$\bA_i=[\bI_{N-i-n}|\bzero_{(N-i-n)\times i}]^T$ with $i\in
\aleph_1^{N-n}$. We further define the following vectors
\begin{equation}
\bpsi^{N-i}_{n,c}=\bpsi_{n,c}^N \cdot \bA_i,~i\in \aleph_1^{N-n},
\quad \bav^{n_2}_{n_1}=[a_{n_1+1},\ldots,a_{n_2}],~n_2 \ge n_1 \ge
0
\end{equation}
where $\bpsi^{N-i}_{n,c}$ is an $(N-i-n)\times 1$ vector and
$\bav_{n_1}^{n_2}$ is an $(n_1-n_2) \times 1$ vector. In particular,
$a_{n_1}^{n_2}$ is defined as $\emptyset$ when $n_1=n_2$.

We next show that the exact value of the integral $I^{(N)}$ in
\eqref{eq:integralI} can be obtained recursively as follows (see
Appendix \ref{proof:lemma2} for the proof).
\begin{Lemma}\label{Lemma:lemma2}
The exact value of the integral $I^{(N)}$ can be obtained by
applying the following recurrence relation
\begin{align}\label{eq:Ialphabetacase}
I^{(N)}=\begin{cases} f^{(N)}_{\sbav^N_0}(b_N)-\mathbb{I}_{\{N\ge
2\}}\displaystyle\sum_{n=0}^{N-2}\frac{(b_N-b_{n+1})^{N-n}}{(N-n)!}I^{(n)},&N\in \aleph_1^Q\\
f^{(N)}_{\sbav^N_0}(b_N)-\displaystyle\sum_{n=0}^{N-2}f^{(N-n)}_{\bpsi^{N}_{n,a_N}}(b_N)I^{(n)},&
N\in \aleph_{Q+1}^{\infty}
\end{cases}
\end{align}
and the initial condition $I^{(0)}=1$.
\end{Lemma}

We now introduce the third integral in the following
\begin{align}
J^{(N)}_{c,d}(\theta):&=\idotsint\limits_{\Upsilon_{c,d}^{(N)}}~e^{-\theta
\xi_N} d\bxi_N \label{eq:integralJ}
\end{align}
where $\theta > 0$, $N\ge 1$, and
$\Upsilon_{c,d}^{(N)}:=\{(\xi_1,\ldots,\xi_N): 0\le \xi_1 \le \ldots
\le \xi_{N};a_i < \xi_i < b_i,i \in \aleph^{N-1}_1;c < \xi_N < d\}$.
Recalling that $c$ is an arbitrary non-negative number satisfying
$a_{N-1} \le c \le b_N$ and $d$ is an arbitrary number satisfying
$0\le c <d$ and $d\ge a_N$, we define the following function
\begin{equation*}
g_{c,d}^{(n)}(\theta)=\begin{cases} \displaystyle I^{(n)}
\Big[\theta^{n-N}e^{-\theta b_{n+1}}-\sum_{i=1}^{N-n} \theta^{-i}
f^{(N-n-i)}_{b_{n+1}\bone_{N-n-i}}(d)e^{-\theta d}\Big],&c \le b_1,~n\in \aleph_0^{N-2} \\
\displaystyle I^{(n)}\sum_{i=1}^{N-n} \theta^{-i}\Big[
f^{(N-n-i)}_{{\bpsi}_{n,c}^{N-i}}(c)e^{-\theta
c}-f^{(N-n-i)}_{{\bpsi}_{n,c}^{N-i}}(d)e^{-\theta d}\Big]
,&c> b_1,~n\in \aleph_0^{s-1} \\
\displaystyle I^{(n)}\Big[\theta^{n-N}e^{-\theta
b_{n+1}}-\sum_{i=1}^{N-n}
\theta^{-i}f^{(N-n-i)}_{b_{n+1}\bone_{N-n-i}}(d)e^{-\theta
d}\Big],&c> b_1,~n\in \aleph_s^{N-2}.
\end{cases}
\end{equation*}
The following lemma shows the exact values of the integral
$J^{(N)}_{c,d}(\theta)$ for two particular pairs of $(c,d)$ (see
Appendix \ref{proof:lemma3} for the proof).
\begin{Lemma}\label{Lemma:lemma3}
For any $\bar{\gamma}_N$ satisfying $a_N \le \bar{\gamma}_N< b_N$,
the exact values of the integrals
$J^{(N)}_{\bar{\gamma}_N,\infty}(\theta)$ and
$J^{(N)}_{a_N,b_N}(\theta)$ are given by
\begin{align}
J^{(N)}_{\bar{\gamma}_N,\infty}(\theta)=&\sum_{i=1}^N\theta^{-i}f^{(N-i)}_{\sbav^{N-i}_0}
(\bar{\gamma}_N)e^{-\theta
\bar{\gamma}_N}-\mathbb{I}_{\{N\ge2\}}\sum_{n=0}^{N-2} g_{\bar{\gamma}_N,\infty}^{(n)}(\theta)\label{eq:jnexpression1} \\
J_{a_{N},b_N}^{(N)}(\theta)=&\sum_{i=1}^N
\theta^{-i}\Big[f^{(N-i)}_{\sbav^{N-i}_0}(a_{N})e^{-\theta
a_{N}}-f^{(N-i)}_{\sbav^{N-i}_0}(b_N)e^{-\theta
 b_N}\Big]
 -\mathbb{I}_{\{N\ge 2\}}\sum_{n=0}^{N-2}g^{(n)}_{a_N,b_N}(\theta).\label{eq:jnexpression3}
\end{align}
\end{Lemma}

We next turn our attention to the evaluation of false-alarm and
miss-detection probabilities.

\subsection{False-Alarm Probability}\label{subsect:falsealarmprobability}

Let $E_N$ denote the event that $\Lambda_N\ge b$ and $a< \Lambda_n
<b$ for $n \in \aleph_1^{N-1}$ with $N\in \aleph_1^{M-1}$, and let
$E_M$ denote the event that  $\Lambda_{M}\ge \gamma$ and $a<
\Lambda_n <b$ for $n \in \aleph_1^{M-1}$. Denote by $P_{H_0}(E_N)$
the probability of the event $E_N$ under $H_0$. Recalling that the
test procedure given in \eqref{eq:h1rule1}-\eqref{eq:continuesample}
is equivalent to that given in
\eqref{eq:h1rule1transform}-\eqref{eq:continuesampletransform}, we
have
\begin{align}\label{eq:ph0en}
P_{H_0}(E_N)= \begin{cases} P_{H_0}(a_i<\xi_i<b_i,i\in
\aleph_1^{N-1};
\xi_{N}\ge b_{N}),&N\in \aleph_1^{M-1}\\
P_{H_0}(a_i < \xi_i < b_i,i \in \aleph_1^{M-1};\xi_M \ge
\bar{\gamma}_M),&N=M.
\end{cases}
\end{align}
Clearly, the false-alarm probability $\alpha_{\text{ssct}}$ can be
written $\alpha_{\text{ssct}}=\sum_{N=1}^{M} P_{H_0}(E_N)$. In the
following proposition, we present the exact false-alarm probability
$\alpha_{\text{ssct}}$.
\begin{Propo}\label{prop:falsealarm}
The false-alarm probability, $\alpha_{\text{ssct}}$, is given by
\begin{align}
\alpha_{\text{ssct}}=\sum_{N=1}^M P_{H_0}(E_N)
\end{align}
where $P_{H_0}(E_N)$ can be computed as
\begin{eqnarray}\label{eq:ph0encases}
\!\!P_{H_0}(E_N)\!=\!\begin{cases} \displaystyle \!
p_N\frac{b_1b_N^{N-2}}{(N-1)!},&N \in \aleph_1^{P+1}
\notag \\
\! p_N\Big[f^{(N-1)}_{\sbav^{N-1}_0}(b_{N-1})\!-\!\mathbb{I}_{\{N\ge
3\}}\displaystyle{\sum_{n=0}^{N-3}}\frac{(b_{N-1}-b_{n+1})^{N-n-1}}{(N-n-1)!}2^{n}e^{\frac{b_{n+1}}{2}}P_{H_0}(E_{n+1})
\Big],&N\in \aleph_{P+2}^{Q+1} \\
\! p_N\Big[f^{(N-1)}_{\sbav^{N-1}_0}(b_{N-1})\!-\!
\displaystyle\sum_{n=0}^{N-3}f^{(N-1-n)}_{\bpsi^{N-1}_{n,a_{N-1}}}(b_{N-1})2^{n}e^{\frac{b_{n+1}}{2}}
P_{H_0}(E_{n+1})\Big],&N\in \aleph_{Q+2}^{M-1} \notag \\
\!2^{-M}J_{\bar{\gamma}_M,\infty}^{(M)}(1/2),&N=M
\end{cases}
\end{eqnarray}
where $p_N:=2^{-(N-1)}e^{-b_N/2}$.
\end{Propo}
\begin{proof} To compute $P_{H_0}(E_N)$, we need to determine the
joint PDF of the RVs $(\xi_1,\ldots,\xi_N)$. Let
$p_{\bv|H_0}(v_1,\ldots,v_N|H_0)$ and
$p_{\bxi|H_0}(\xi_1,\ldots,\xi_N|H_0)$ denote the joint PDFs of the
RVs $(v_1,\ldots,v_N)$ and the RVs $(\xi_1,\ldots,\xi_N)$ under
$H_0$, respectively. Recalling that $v_i$ is an exponential RV
distributed according to \eqref{eq:pdfh0}, we can write the joint
PDF of the RVs $(v_1,\ldots,v_N)$ as
$p_{\bv|H_0}(v_1,\ldots,v_N|H_0)=2^{-N}e^{-\sum_{i=1}^N v_i/2}$. Due
to $\xi_N=\sum_{i=1}^N v_i$, we have
$v_1=\xi_1,v_2=\xi_2-\xi_1,\ldots,v_N=\xi_N-\xi_{N-1}$, which yields
\begin{equation}\label{eq:pdfu}
p_{\bxi|H_0}(\xi_1,\xi_2,\ldots,\xi_N|H_0)=p_{\bv|H_0}(\xi_1,\xi_2-\xi_1,\ldots,\xi_N-\xi_{N-1}|H_0)=2^{-N}e^{-\xi_N/2},~
\end{equation}
where $\xi_0:=0 \le \xi_1\le \cdots \le \xi_N$. According to
\eqref{eq:ph0en} and the definition of
$\Upsilon_{b_N,\infty}^{(N)}$, we have
\begin{equation}\label{eq:ph0endetail}
P_{H_0}(E_N)=P_{H_0}\Big((\xi_1,\xi_2,\ldots,\xi_N)\in
\Upsilon_{b_N,\infty}^{(N)}\Big)
=\idotsint\limits_{\Upsilon_{b_N,\infty}^{(N)}}2^{-N}e^{-\xi_N/2}d\bxi_N.
\end{equation}
Generally speaking, because each variable $\xi_i$ is lower-bounded
by the maximum of $a_i$ and $\xi_{i-1}$, and is upper-bounded by the
minimum of $b_i$ and $\xi_{i+1}$, a direct evaluation is highly
complex due to numerous possibilities of upper- and lower-limits of
$(\xi_1,\xi_2,\ldots,\xi_N)$
\cite{IEEEexample:article_exacttruncatedsequential}.

Nevertheless, in the case of $N\in \aleph_1^{P+1}$, the parameters
$a_i$ for $i\in \aleph_1^{N-1}$ are all zeros by the definition of
the parameter $P$. This implies that $\xi_i$ is only lower-bounded
by $\xi_{i-1}$ for $i\in \aleph_1^{N-1}$, and accordingly the
upper-bound of $\xi_i$ can be readily identified as $b_i$ for $i\in
\aleph_1^{N-1}$ \cite{IEEEexample:article_exacttruncatedsequential}.
Using \cite[Eqs. (16) and
(17)]{IEEEexample:article_exacttruncatedsequential} and the fact
that $\{b_i\}_{i=1}^{\infty}$ is an arithmetic sequence, we obtain
$P_{H_0}(E_N)$ as follows
\begin{align}\label{eq:prh0en1}
P_{H_0}(E_N)= \int_{\xi_0}^{b_{1}}d\xi_1
\int_{\xi_1}^{b_2}d\xi_2\cdots\int_{\xi_{N-2}}^{b_{N-1}}d\xi_{N-1}
\int_{b_N}^{\infty} 2^{-N}e^{-\xi_N/2} d\xi_N
=\frac{b_1b_N^{N-2}e^{-b_N/2}}{2^{N-1}(N-1)!}
\end{align}
with $N\in \aleph_1^{P+1}$.

We now consider the case of $N\in \aleph_{P+2}^{M-1}$. Since
$\xi_N\in [b_N,\infty)$ and $b_N>b_{N-1}$, the upper-limit of
$\xi_{N-1}$ is actually $b_{N-1}$ irrespective of $\xi_N$. Hence,
from \eqref{eq:integralI}, we can write \eqref{eq:ph0endetail} as
\begin{equation}\label{eq:prh0en2}
P_{H_0}(E_N)=
\int_{b_N}^{\infty}2^{-N}e^{-\xi_N/2}d\xi_N\cdot\idotsint\limits_{\Omega^{(N-1)}}
d\bxi_{N-1}=2^{-(N-1)}e^{-b_N/2}I^{(N-1)}
\end{equation}
with $N\in \aleph_{P+2}^{M-1}$. By applying \eqref{eq:integralI},
$P_{H_0}(E_N)$ for $N\in \aleph_{P+2}^{Q+1}$ and $N\in
\aleph_{Q+2}^{M-1}$ in \eqref{eq:ph0encases} can be readily
obtained.

We next compute $P(E_{M}|H_0)$. Since $\bar\gamma_M \in (a_M,b_M)$,
the upper-limit of $\xi_{M-1}$ depends on both $\xi_{M}$ and
$b_{M-1}$. Thus, the integrals $\int d\xi_M$ and $\idotsint\limits
d\bxi_{M-1}$ are not separable. It is clear from Lemma
\ref{Lemma:lemma3} that $P(E_M|H_0)$ can be obtained from
$J_{\bar{\gamma}_M,\infty}^{(M)}(\theta)$ by setting $\theta=1/2$
and $N=M$ in \eqref{eq:jnexpression1}. Hence, we have
\begin{equation}\label{eq:prh0em}
P_{H_0}(E_M)=P_{H_0}\big((\xi_1,\ldots,\xi_M)\in
\Upsilon_{\bar{\gamma}_M,\infty}^{(M)}\big)=
\idotsint\limits_{\Upsilon_{\bar{\gamma}_M,\infty}^{(M)}}
2^{-M}e^{-\xi_M/2}d\bxi_M=2^{-M}J_{\bar{\gamma}_M,\infty}^{(M)}
\big(1/2\big).
\end{equation}
Thanks to \eqref{eq:prh0en1}, \eqref{eq:prh0en2}, and
\eqref{eq:prh0em}, we can conclude the proof.
\end{proof}

\begin{Remark}
The problem of evaluating \eqref{eq:ph0endetail} resembles the exact
OC evaluation problem in truncated sequential life tests involving
the exponential distribution
\cite{IEEEexample:article_exacttruncatedsequential}. Exact OC for
truncated sequential life tests in the exponential case has been
solved by Woodall and Kurkjian
\cite{IEEEexample:article_exacttruncatedsequential}. However, the
Woodall-Kurkjian approach
\cite{IEEEexample:article_exacttruncatedsequential} is not
applicable to evaluate the ASN \cite{Aroian68:directmethod}. More
importantly, it cannot be used to evaluate \eqref{eq:prh0em}
directly. In the preceding proof, we propose a new approach to
derive the exact false-alarm probability $\alpha_{\text{ssct}}$. It
is worth mentioning that with slight modifications, our approach is
also applicable to evaluate the ASN for truncated sequential life
tests in the exponential case.
\end{Remark}

\subsection{Miss-Detection Probability}\label{subsect:missdetectionprobability}

We now turn to the evaluation of the miss-detection probability,
$\beta_{\text{ssct}}$. In order to evaluate $\beta_{\text{ssct}}$,
we need to obtain $p_{H_1}(v_i)$. Since acquiring perfect knowledge
on instantaneous $\lambda_i$ or the exact distribution of
$\lambda_i$ may not be feasible in practice, evaluating
$p_{H_1}(v_i)$ is typically difficult except for the case where the
primary signals are constant-modulus, i.e., $|s_i|^2=\sigma_s^2$. We
next reason that at a relatively low detection SNR level, the
miss-detection probability obtained by using constant-modulus
primary signals can be used to well approximate the actual
$\beta_{\text{ssct}}$. Our arguments are primarily based on the
following two properties of the SSCT.

The first property shows that as $N$ approaches infinity, the
distribution of the test statistic $\xi_N$ in the SSCT converges to
a normal distribution that is independent of a specific choice of
$\lambda_1,\lambda_2,\ldots,\lambda_N$.
\begin{Property}\label{property1}
The statistical distribution of $\xi_N$ converges to a normal
distribution given by
\[
\xi_N \sim \begin{cases} \mathcal{N}(2N,4N), & \text{under}~H_0, \\
\mathcal{N}\big(2N(1+{\tt{SNR}}_m), 4N(1+2{\tt{SNR}}_m)\big), &
\text{under}~H_1,
\end{cases}
\]
as $N$ approaches infinity.
\end{Property}

The property can be readily proved by using the CLT
\cite{surveyjohnson:sequential}. However, unlike energy detection,
this property alone is not sufficient to explain that the
constant-modulus assumption is valid in approximating
$\beta_{\text{ssct}}$. This is because each $\xi_N$ for
$N=1,\ldots,M$ including small values of $N$, may potentially affect
the value of $\beta_{\text{ssct}}$. To complete our argument, we
first present the following definitions. Let $\tilde {\xi}_i$ denote
the test statistic using the constant-modulus assumption, i.e.,
$|s_i|^2=\sigma_s^2$. Define $\varrho_N:=b_N$ for $N\in
\aleph_1^{M-1}$ and $\varrho_M:=\bar \gamma_M$ for $N=M$. Let $F_N$
and $\tilde{F}_N$ denote the events that $\xi_N \ge \varrho_N$, and
$a_i < \xi_i < b_i$ for $i\in \aleph_1^M$, and $\tilde \xi_N \ge
\varrho_N$ and $a_i < \tilde \xi_i < b_i$ for $i\in \aleph_1^M$,
respectively. Let $P_{H_1}(F_N)$ and $P_{H_1}(\tilde F_N)$ denote
the probabilities of the events $F_N$ and $\tilde F_N$ under $H_1$.
Let $\tilde{\beta}_{\text{ssct}}$ denote the miss-detection
probability obtained by assuming constant-modulus signals with
average symbol energy $\sigma_s^2$, i.e., $|s_i|^2=\sigma_s^2$. As
clear from their definitions, we have
$\beta_{\text{ssct}}=\sum_{N=1}^M P_{H_1}(F_N)$ and $\tilde
\beta_{\text{ssct}}=\sum_{N=1}^M P_{H_1}(\tilde F_N)$.

Let $\mathcal{A}_{N}^{l_N}$ denote the event that $a_i < \xi_i <
b_i,~i\in \aleph_1^{l_N}$ for some integer $l_N\in \aleph_1^N$, and
let $\tilde {\mathcal{A}}_N^{l_N}$ denote its counterpart for the
constant-modulus case. Let $\mathcal{B}_N^{l_N}$ denote the event
that $\xi_N\ge \varrho_N$ and $a_i < \xi_i < b_i,~i\in
\aleph_{l_N+1}^N$ and let $\tilde {\mathcal{B}}_N^{l_N}$ denote its
counterpart in the constant-modulus case. We now present the second
property of the SSCT (see Appendix \ref{proof:property2} for the
proof).

\begin{Property}\label{property2}
Let $\epsilon$ an arbitrary positive number. If for each $N$, there
exists a positive integer $l_N \in \aleph_1^N$ such that
$P_{H_1}(\mathcal{A}_N^{l_N})\ge 1-\epsilon/(3M)$,
$P_{H_1}(\tilde{\mathcal{A}}_N^{l_N})\ge 1-\epsilon/(3M)$, and
$|P_{H_1}(\mathcal{B}_N^{l_N})-P_{H_1}(\tilde{\mathcal{B}}_N^{l_N})|<\epsilon/(3M)$,
then
\[
|\beta_{\text{ssct}}-\tilde \beta_{\text{ssct}}|\le \epsilon
\]
where $l_N$ depends on the values of $N$ and $\epsilon$.
\end{Property}

Relying on these two properties, we sketch our arguments as follows.
To achieve high detection accuracy at a low SNR level, the ASN and
$M$ are typically quite large. When the sample index $N$ is
relatively small, it is highly likely that the test statistics
$\xi_N$ and $\tilde{\xi}_N$ do not cross either of two boundaries.
In such a situation, there exists some integer $l_N$ such that
$P_{H_1}(\mathcal{A}_N^{l_N})$ and
$P_{H_1}(\tilde{\mathcal{A}}_N^{l_N})$ are fairly close to $1$
whereas $P_{H_1}(\mathcal{B}_N^{l_N})$ and
$P_{H_1}(\tilde{\mathcal{B}}_N^{l_N})$ are fairly close to $0$.
Hence, the conditions in Property \ref{property2} can be easily
satisfied. On the other hand, when $N$ is relatively large, one can
find a sufficiently large $l_N$ such that
$P_{H_1}(\mathcal{A}_N^{l_N})$ and
$P_{H_1}(\tilde{\mathcal{A}}_N^{l_N})$ are fairly close to one while
$|P_{H_1}(\mathcal{B}_N^{l_N})-P_{H_1}(\tilde{\mathcal{B}}_N^{l_N})|$
are sufficiently small due to Property \ref{property1} guaranteed by
the CLT. Collectively, at a low detection SNR level, $\tilde
{\beta}_{\text{ssct}}$ evaluated under the constant-modulus
assumption is a close approximation of $\beta_{\text{ssct}}$.
Therefore, we will focus on the case in which all $\lambda_i$'s are
equal to a constant
$\lambda:=2|h|^2\sigma_s^2/\sigma_w^2=2{\tt{SNR}}_m$.

Recall that under $H_1$, $v_i$ is a non-central chi-square RV, whose
PDF involves the zeroth-order modified Bessel function of the first
kind as given in \eqref{eq:pdfh1}. This makes it infeasible to
evaluate $\tilde{\beta}_{\text{ssct}}$ by applying the computational
approach used in the false-alarm probability case. To obtain
$\tilde{\beta}_{\text{ssct}}$, we resort to a numerical integration
algorithm proposed in \cite{pollockgolhar:technicalreport}.

Defining $u_i = v_i -\bar \Delta$, we rewrite $\bar{\Lambda}_N$ in
\eqref{eq:lambdabar} as $\bar{\Lambda}_N=\sum_{i=1}^N u_i$. Clearly,
we can write the PDF of $u_i$ under $H_1$ as
\begin{equation}\label{eq:ph1uilambda}
p_{H_1}(u_i) = \frac{1}{2}e^{-(u_i + \bar
\Delta+\lambda)}I_0\Big(\sqrt{\lambda(u_i+\bar
\Delta)}\Big),u_i>-\bar\Delta.
\end{equation}
Recall that $M$ is the maximum number of samples to observe. Denote
$\bar\Lambda_{M-k}$ by $t_k$. Let $G_k(t_k)$ denote the conditional
miss-detection probability of the SSCT conditioning on that the
first $(M - k)$ samples have been observed, the present value
$t_k=\bar \Lambda_{M-k}$, and the test statistic has not crossed
either boundary in the previous $(M-k-1)$ samples. If $\bar a < t_k
< \bar b$, an additional sample (the $(M-k+1)$th sample) is needed.
Let $u$ be the next observed value of $u_i$. The conditional
probability $G_k(t_k|u)$ can be written as
\begin{align}\label{eq:rhokrecursive}
G_k(t_k|u) = \begin{cases} 0  &\textrm {if}~ u > \bar
b-t_k \\
                          1  &\textrm {if}~  u < \bar a-t_k \\
                          G_{k-1}(t_{k}+u) &\textrm{if}~\bar a-t_k\le u \le \bar b-t_k
                          .
        \end{cases}
\end{align}
Using \eqref{eq:rhokrecursive}, we can compute $G_k(t_k)$ as
\begin{align}
\label{eqn:gktk} &G_k(t_k)=\int_{-\infty}^{\bar a - t_{k}}
p_{H_1}(u) du
   + \int_{\bar a - t_{k}}^{\bar b - t_{k}} G_{k-1}(t_{k}+u) p_{H_1}(u) du
\end{align}
for $k = 1, \cdots, M$ with the following initial condition:
\begin{eqnarray}
\label{eqn:g0t0}
G_0(t_0) = \begin{cases} 0, & \textrm{ if }  t_0 \ge \bar \gamma \cr
                          1, & \textrm{ otherwise.}
\end{cases}
\end{eqnarray}
Employing the above backward recursion process, we can obtain
$G_M(0)$, which is equal to the miss-detection probability,
$\tilde{\beta}_{\text{ssct}}$.

\section{Evaluation of the Average Sample Number}\label{section:averagesensingtime}

Roughly speaking, the false-alarm and miss-detection probabilities,
and the ASN are three principal performance benchmarks for the
sequential sensing scheme. In the preceding section, we have only
concerned ourselves with the error probability performance while in
this section, we show how to evaluate the ASN.

Let $N_s$ denote the number of samples required to yield a
decision. Clearly, $N_s$ is a RV in the SSCT, and its mean value
is the ASN, which can be written as
\begin{equation}\label{eq:asn1}
E(N_s)=E_{H_0}(N_s)P_{H_0}(H_0)+E_{H_1}(N_s)P_{H_1}(H_1)
\end{equation}
where $E_{H_i}(N_s)$ denotes the ASN conditional on $H_i$, and
$P_{H_i}(H_i)$ denotes {\it a priori} probability of hypothesis
$H_i$, for $i=0,1$.

According to \eqref{eq:h1rule1}--\eqref{eq:continuesample}, we have
$1\le N_s\le M$. Hence, we can express $E_{H_i}(N_s)$ as
\begin{equation}\label{eq:ehins}
E_{H_i}(N_s)=\sum_{N=1}^M NP_{H_i}(N_s=N),~i=0,1
\end{equation}
where $P_{H_i}(N_s=N)$ is the conditional probability that the
detector makes a decision at the $N$th sample under $H_i$. We now
need to determine $P_{H_i}(N_s=N)$. Let ${\mathcal{C}}_N$ denote the
event that the test statistics $(\xi_1,\xi_2,\ldots,\xi_N)$ do not
cross either the upper or lower boundary before or at the $N$th
sample, i.e., ${\mathcal{C}}_N=\{(\xi_1,\xi_2,\ldots,\xi_N)\in
\Upsilon_{a_{N},b_{N}}^{(N)}\}$ for $N \in \aleph_1^{M}$. For
notional convenience, let us define ${\mathcal{C}}_0$ as a universe
set. Hence, we have $P({\mathcal{C}}_0)=1$. It implies from
\eqref{eq:h1rule1transform}-\eqref{eq:continuesampletransform} that
$P_{H_i}(N_s=N),~i=0,1$ can be obtained as
\begin{align}
P_{H_i}(N_s=N)&\overset{(a)}{=}P_{H_i}\big({\mathcal{C}}_{N-1}\big)-P_{H_i}\big({\mathcal{C}}_{N}\big),~N\in
\aleph_1^{M-1}
\label{eq:phinsn} \\
P_{H_i}(N_s=M)&\overset{(b)}{=}P_{H_i}\big({\mathcal{C}}_{M-1}\big),\quad\quad
\quad \quad \quad N=M\label{eq:phinsnM}
\end{align}
where the two terms on the right-hand side (RHS) of the equality
$(a)$ are the probabilities of the events that under $H_i$, the test
statistic does not cross either of two boundaries before or at the
$(N-1)$th sample and the $N$th sample for $N\in \aleph_1^{M-1}$,
respectively, and the term on the RHS of the equality $(b)$ denotes
the probability of the event that under $H_i$, the test statistic
does not cross either boundary before or at the $(M-1)$th sample.

Substituting (\ref{eq:phinsn}) and (\ref{eq:phinsnM}) in
(\ref{eq:ehins}), we can rewrite (\ref{eq:ehins}) as
\begin{align}
E_{H_i}(N_s) =\sum_{N=1}^{M-1} N
\Big(P_{H_i}\big({\mathcal{C}}_{N-1}\big)
-P_{H_i}\big({\mathcal{C}}_N\big)\Big)+M
P_{H_i}\big({\mathcal{C}}_{M-1}\big)= 1 + \sum_{N=1}^{M-1}
P_{H_i}\Big({\mathcal{C}}_N\Big).\label{eq:ehinsrw}
\end{align}

According to Lemma \ref{Lemma:lemma3}, \eqref{eq:pdfu},
\eqref{eq:phinsn}, and \eqref{eq:phinsnM}, we have
\begin{equation}\label{eq:ph0set}
P_{H_0}\big({\mathcal{C}}_N\big)=2^{-N}J^{(N)}_{a_N,b_N}(1/2),~N\in
\aleph_1^{M-1}.
\end{equation}
Hence, from \eqref{eq:ehinsrw}, we have
\begin{equation}\label{eq:eh0nsexact}
E_{H_0}(N_s)=1+\sum_{N=1}^{M-1}2^{-N}J^{(N)}_{a_N,b_N}(1/2).
\end{equation}
%

We next need to evaluate $P_{H_1}({\mathcal{C}}_N)$. Instead of
evaluating $P_{H_1}({\mathcal{C}}_N)$ directly, we compute
$P_{H_1}({\mathcal{C}}^c_N)$ by performing a procedure similar to
the one in calculating the miss-detection probability. Note that
${\mathcal{C}}_N^c$ denotes the event that under $H_1$, the test
procedure given in
\eqref{eq:h1rule1transform}-\eqref{eq:continuesampletransform}
terminates before or at the $N$th sample, i.e., the test statistic
crosses either the upper or lower boundary before or at the $N$th
sample. According to \eqref{eqn:g0t0}, $G_k(t_k)$ also depends on
$\bar \gamma$. With a slight abuse of notation, we rewrite
$G_k(t_k)$ as $G_k(t_k, \bar \gamma)$. Let $V_N$ denote the event
that the test statistic crosses the lower boundary before or at the
$N$th sample under $H_1$, and $U_N$ denote the event that the test
statistic {\em does not} cross the upper-boundary before or at the
$N$th sample under $H_1$. It is clear to see $P_{H_1}(V_N) = G_N(0,
\bar a)$ and $P_{H_1}(U_N) = G_N(0, \bar b)$. Obviously,
$P_{H_1}(\mathcal{C}_N^c)$ can be written as
\begin{align}
P_{H_1}\big({\mathcal{C}}_N^c\big) = P_{H_1} (V_N) + (1 -
P_{H_1}(U_N))=G_N(0, \bar a) + 1 - G_N(0, \bar b)
  \label{eqn:ph1sn3}
\end{align}
where $G_N(t, \bar a)$ and  $G_N(t, \bar b)$ can be obtained by
applying \eqref{eqn:gktk} recursively. Hence, we have
$P_{H_1}({\mathcal{C}}_N)=G_N(0, \bar b) - G_N(0, \bar a)$.
According to \eqref{eq:ehinsrw}, we have
\begin{align}\label{eq:enh1nsnumerical}
E_{H_1}(N_s) = 1 +  \sum_{N=1}^{M-1} (G_N(0, \bar b) - G_N(0, \bar
a)).
\end{align}
In the following proposition, we present the ASN of the SSCT.
\begin{Propo}\label{propo:averagelengthavageexact}
The ASN of the SSCT can be obtained as
\begin{align}
E(N_s)=&~ P_{H_0}(H_0)\Big(1+ \sum_{N=1}^{M-1}
2^{-N}J^{(N)}_{a_N,b_N}(1/2)\Big)+P_{H_1}(H_1)\Big(1+\sum_{N=1}^{M-1}
(G_N(0, \bar b) - G_N(0, \bar a))\Big).
\end{align}
\end{Propo}
\begin{proof}
The proof follows immediately from \eqref{eq:asn1},
\eqref{eq:eh0nsexact}, and \eqref{eq:enh1nsnumerical}.
\end{proof}

\section{Simulations}\label{section:simulation}

In this section, we provide several examples to illustrate the
effectiveness of the SSCT. In all simulation examples, we select
$\bar\Delta$ to be $2+{\tt{SNR}}_m$, which ensures
$E_{H_0}(\Lambda_N)=-E_{H_1}(\Lambda_N)$. In the first three test
examples, the truncated sample size $M$ is selected to be the
minimum sample number $M_{\text{ed}}^{\min}$ required by energy
detection to achieve specified false-alarm and miss-detection
probabilities, and $\bar a$ is always chosen to be $-\bar b$. All
the test examples assume that $|h|$ is equal to one, and adopt
constant-modulus quadrature phase shift-keying (QPSK) signals except
for Test Example 2, in which the modulation formats of the primary
signals are explicitly stated. Following the conventional
terminology in sequential detection, we define the efficiency of the
SSCT as $\eta:=1-\text{ASN}/M_{\text{ed}}^{\min}$.

{\it Test Example 1 (The SSCT Versus Energy Detection):} Table
\ref{table1:simulation} compares the SSCT with energy detection in
terms of false-alarm and miss-detection probabilities and the ASN
for different ${\tt{SNR}}_m$. The truncated sizes corresponding to
${\tt{SNR}}_m=0$, $-5$, $-10$, and $-15$ dB, are selected to be the
minimum sample sizes required by energy detection to achieve target
$(\bar\alpha_{\text{ssct}},\bar\beta_{\text{ssct}})=(0.01,0.01)$,
$(0.05,0.05)$, $(0.1,0.1)$, and $(0.15,0.15)$, respectively. The
parameters $\bar b,~\bar \Delta$ and $\bar \gamma$ are given in the
table. It is shown in Table \ref{table1:simulation} that compared
with energy detection, the SSCT can achieve about $29\%\sim 35\%$
savings in the average sensing time while maintaining a comparable
detection performance. It can be also observed from the table that,
as indicated by using an abbreviation, Numerical, in the
parenthesis, the false-alarm probabilities computed from the exact
result \eqref{eq:ph0encases} and the miss-detection probabilities
obtained by the numerical integration algorithm match well with
those obtained by Monte Carlo simulations.

{\it Test Example 2 (Detection Performance Without Knowing the
Modulation Format of the Primary Signals)}: In this example, the
primary signals are modulated by using a square 64-quadrature
amplitude modulation (QAM) with
$\sigma_s^2=10^{{{\tt{SNR}}_m}/10}\sigma_w^2$. Table
\ref{table2:simulation} compares with the miss-detection
probabilities and the ASN between the SSCT and energy detection for
${\tt {SNR}}_m=0$, $-5$, $-10$, and $-15$ dB. The parameters $\bar
a$, $\bar b$, $\bar \gamma$, $\bar \Delta$, and $M$ in the SSCT, are
determined by using constant-modulus QPSK signals with the same
average symbol energy $\sigma_s^2=10^{{\text{SNR}_m}/10}\sigma_w^2$,
and these design parameters are used in the SSCT to detect the
64-QAM primary signals. That is, the SSCT does not have the
knowledge of the modulation format of the primary signals. When
evaluating the miss-detection probability and the ASN of the SSCT in
the 64-QAM case, we use equiprobable prior distributions of $s_i$ to
obtain $p_{H_1}(u_i)$. As can be seen from the table, the
miss-detection probabilities obtained in the 64-QAM case are fairly
close to the ones obtained in the QPSK case except for the case of
${\tt {SNR}}_m=0$ dB, where $M=40$. This is because $M$ is not large
enough to neglect errors caused by using the CLT approximation.
However, energy detection and the SSCT suffer from a similar amount
of approximation error when evaluating the miss-detection
probability.

{\it Test Example 3 (Mismatch between ${\tt{SNR}}_o$ and
${\tt{SNR}}_m$)}: Table \ref{table3:simulation} lists the
false-alarm and miss-detection probabilities and ASN when
${\tt{SNR}}_o=-12$, $-13$, $-14$ dB, and ${\tt{SNR}}_{m}=-15$ dB.
The parameters for the SSCT and energy detection are determined to
ensure that a target pair
$(\alpha_{\text{ssct}},\beta_{\text{ssct}})$ satisfies $(0.15,0.15)$
at ${\tt{SNR}}_m=-15$ dB. It can be seen from the table that, as
${\tt{SNR}}_o$ increases, the miss-detection probability decreases
while the false-alarm probability remains unchanged. All obtained
$(\alpha_{\text{ssct}},\beta_{\text{ssct}})$ pairs satisfy the
target false-alarm and miss-detection probability requirement.
Interestingly, the efficiency of the SSCT increases to $46\%$ from
$29\%$ even though the miss-detection probabilities of the SSCT are
larger than those in energy detection. It implies from this example
that the SSCT offers more flexibility in striking the tradeoff
between sensing time and detection performance than energy
detection.

{\it Test Example 4 (Impacts of Truncated Size $M$ on the Efficiency
of the SSCT)}: Let $T_p$ denote the probability of the event that
the SSCT ends at the $M$th sample (truncated at the $M$th sample).
Table \ref{table4:simulation} lists $T_p$, the ASN, and the
efficiency of the SSCT for various selected combinations of $M$,
$\bar a$, $\bar b$, and $\bar \gamma$ at ${\tt{SNR}}_m=-5$ dB to
achieve target
$(\bar\alpha_{\text{ssct}},\bar\beta_{\text{ssct}})=(0.055,0.046)$.
The results shown in the table are obtained by using Monte Carlo
simulation. To achieve roughly the same false-alarm and
miss-detection probabilities, the sample size for energy detection
is chosen to be $140$. As can be seen from this table, the
efficiency of the SSCT increases as the truncated size $M$ increases
but the pace of the improvement is diminishing. Table
\ref{table4:simulation} also lists the efficiency of the
non-truncated SPRT based sensing scheme presented in Section
\ref{subsubsection:sprt}. It is clear from the table that as the
truncated size $M$ increases, the efficiency of the SSCT comes
fairly close to the one achieved by the {\it non-truncated} SPRT
based sensing scheme.

\section{Conclusion}\label{section:conclusion}

In cognitive radio networks, stringent requirements on the secondary
and opportunistic access to licensed spectrum, necessitate the need
to develop a spectrum sensing scheme that is able to quickly detect
weak primary signals with high accuracy in a non-coherent fashion.
Motivated by this, we have proposed a sequential sensing scheme that
possesses several desirable features suitable for cognitive radio
networks. To efficiently and accurately obtain major performance
benchmarks of our sensing scheme, we have derived an exact result
for the false-alarm probability and have applied a numerical
integration algorithm to compute the miss-detection probability and
the ASN.

There are several potential extensions of this work that deserve
further exploration. First, our approach to determining design
parameters such as thresholds and the truncated size follows the
original Wald's approach in the sense that the cost of observations
as well as the cost of the false-alarm and miss-detection events
have not been considered. A Bayesian formulation of the SSCT can be
an interesting extension. Second, this work assumes perfect
knowledge on noise power at the SU, which perhaps is difficult to
acquire in practice. The effect of noise power uncertainty on the
SSCT is worth investigating. Third, another extension of this work
is to study sensing-throughput tradeoffs for the SSCT.

\appendices

\section{Proof of Lemma \ref{Lemma:lemma1}}\label{proof:lemma1}

We prove the lemma by induction. It is obvious from \eqref{eq:giu}
that \eqref{eq:fchikalternative} holds for $k=1$. Now suppose that
\eqref{eq:fchikalternative} and \eqref{eq:recursive1} hold for the
case of $k-1$. By definition and the induction assumption for
$k-1$, we have
\begin{align}
f_{\bchi_k}^{(k)}(\xi)&=\int_{\chi_k}^\xi \left(\sum_{i=0}^{k-2}
\frac{f_i^{(k-1)}}{(k-1-i)!}(\xi_{k-1}-\chi_{i+1})^{k-1-i}+f_{k-1}^{(k-1)}\right)d\xi_{k-1}\notag
\\ &=\sum_{i=0}^{k-2}
\frac{f_{i}^{(k-1)}}{(k-i)!}(\xi_{k-1}-\chi_{i+1})^{k-i}\Big|_{\chi_k}^{\xi}+f_{k-1}^{(k-1)}\cdot
(\xi-\chi_k)\notag
\\
&=\sum_{i=0}^{k-1}\frac{f_i^{(k-1)}}{(k-i)!}(\xi-\chi_{i+1})^{k-i}-\sum_{i=0}^{k-2}
\frac{f_i^{(k-1)}}{(k-i)!}(\chi_k-\chi_{i+1})^{k-i}.\label{eq:fchikproof}
\end{align}
Clearly, comparing \eqref{eq:fchikalternative} with
\eqref{eq:fchikproof}, we can readily conclude the recurrence
relation given in \eqref{eq:recursive1}. In particular, when
$\chi_1=\chi_2=\ldots=\chi_k$, all coefficients $f_i^{(k)}$ except
$f_0^{(k)}$ are zeros and hence \eqref{eq:specialcasefk} follows
immediately.

Since the differential property can be proved in a straightforward
manner, we omit the proof. We next prove the scaling property by
induction. When $k=1$, we have
\[
f_{t{\bchi_1}}^{(1)}(t\xi)=\int_{t\chi_1}^{t\xi}
d\xi_1=t(\xi-\chi_1)=tf_{\bchi_1}^{(1)}(\xi).
\]
Hence, the scaling property holds for $k=1$. We now suppose that the
property holds for $k=n-1$. Applying the induction assumption and a
substitution $t\xi_n=u$, we can rewrite the integral
$f_{\bchi_n}^{(n)}(\xi)$ as
\[
f_{\bchi_n}^{(n)}(\xi)=\int_{\chi_n}^{\xi}
f_{\bchi_{n-1}}^{(n-1)}(\xi_n)d\xi_n=\int_{\chi_n}^{\xi}
t^{-(n-1)}f_{t\bchi_{n-1}}^{(n-1)}(t\xi_n)d\xi_n=t^{-n}\int_{t\chi_n}^{t\xi}
f_{t\bchi_{n-1}}^{(n-1)}(u)du=t^{-n}f_{t\bchi_{n}}^{(n)}(t\xi).
\]
Hence, the scaling property holds for $k=n$. This concludes the
proof. The shift property can be proved in a similar manner and thus
the proof is omitted. \hfill $\blacksquare$

\section{Proof of Lemma \ref{Lemma:lemma2}}\label{proof:lemma2}

\begin{figure}
\centering \subfigure[Possible Paths from $\Omega^{(N)}~(N\le Q)$.]{
\psfrag{a1}{\footnotesize{$a_1$}}\psfrag{a2}{\footnotesize{${a_2}$}}\psfrag{an1}{\footnotesize{$a_{N-1}$}}
\psfrag{an}{\footnotesize{$a_N$}} \psfrag{aQ}{\footnotesize{$a_Q$}}
\psfrag{aQ1}{\footnotesize{$a_{Q+1}$}}
\psfrag{ap}{\footnotesize{$a_P$}} \psfrag{b1}{\footnotesize{$b_1$}}
\psfrag{b2}{\footnotesize{$b_2$}}\psfrag{bn1}{\footnotesize{$b_{N-1}$}}\psfrag{bn}{\footnotesize{$b_{N}$}}
\psfrag{a}{\footnotesize{$a$}} \psfrag{b}{\footnotesize{$b$}}
\psfrag{x}{\footnotesize{$x$}} \psfrag{y}{\footnotesize{$y$}}
\psfrag{xi1}{\footnotesize{$\xi_1$}}\psfrag{xi2}{\footnotesize{$\xi_2$}}
\psfrag{xiN1}{\footnotesize{$\xi_{N-1}$}}\psfrag{xiN}{\footnotesize{$\xi_N$}}
\psfrag{Numberofsamples} {\footnotesize{$\tt{Sample~Index}$}}
\psfrag{N}{\footnotesize{$N$}}
\epsfig{file=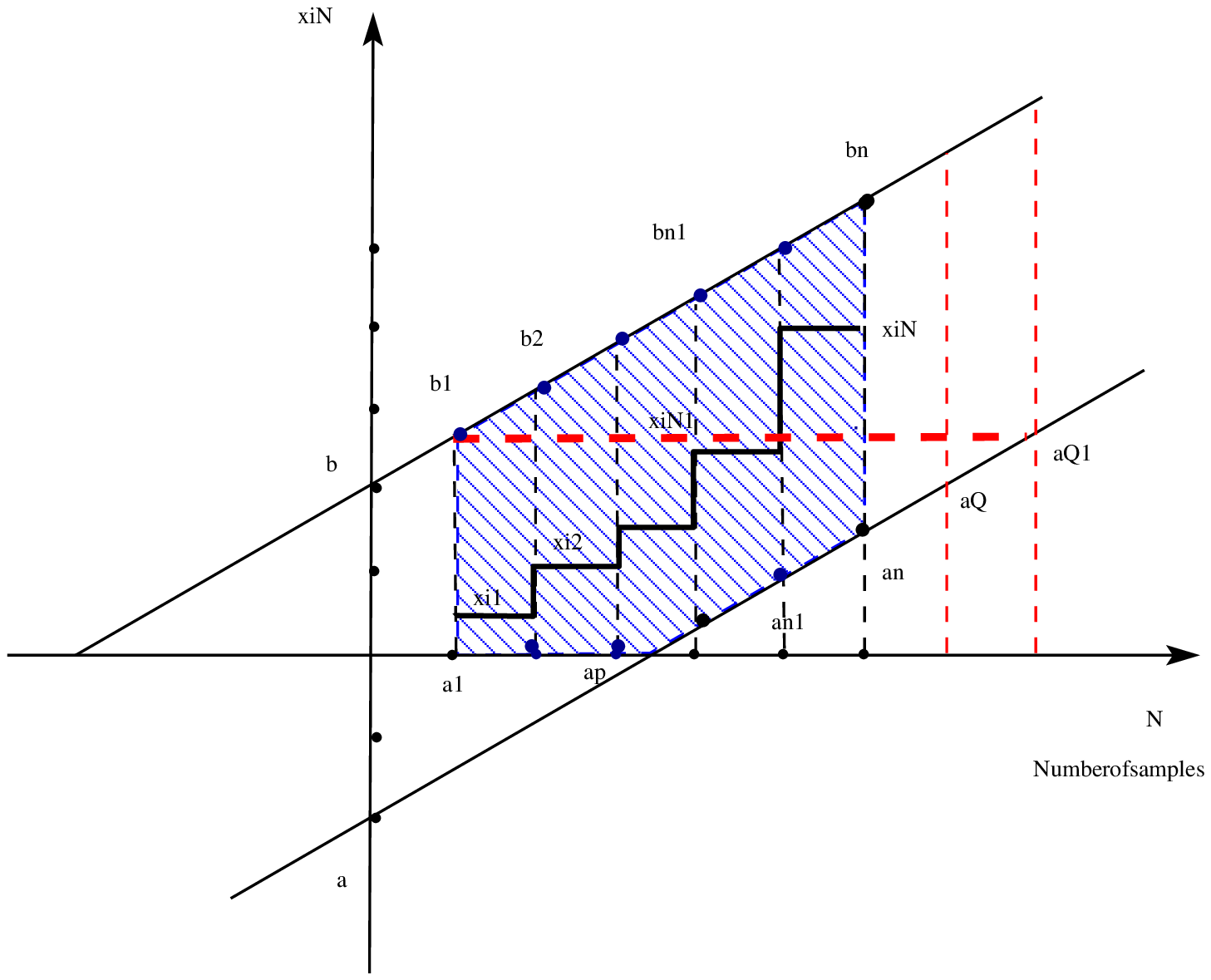,width=0.43\textwidth} \label{fig:rn}}
\subfigure[Possible Paths from $\Pi_{\bav^N_0}^{(N)}$.]{
\psfrag{a1}{\footnotesize{$a_1$}}\psfrag{a2}{\footnotesize{${a_2}$}}\psfrag{an1}{\footnotesize{$a_{N-1}$}}
\psfrag{an}{\footnotesize{$a_N$}} \psfrag{b1}{\footnotesize{$b_1$}}
\psfrag{b2}{\footnotesize{$b_2$}}\psfrag{bn1}{\footnotesize{$b_{N-1}$}}\psfrag{bn}{\footnotesize{$b_{N}$}}
\psfrag{a}{\footnotesize{$a$}} \psfrag{b}{\footnotesize{$b$}}
\psfrag{x}{\footnotesize{$N$}}
\psfrag{Numberofsamples}{\footnotesize{$\tt{Sample~Index}$}}
\psfrag{y}{\footnotesize{$\xi_N$}}\psfrag{ap}{\footnotesize{$a_P$}}
\psfrag{xi1}{\footnotesize{$\xi_1$}}\psfrag{xi2}{\footnotesize{$\xi_2$}}
\psfrag{xiN1}{\footnotesize{$\xi_{N-1}$}}\psfrag{xiN}{\footnotesize{$\xi_N$}}
\epsfig{file=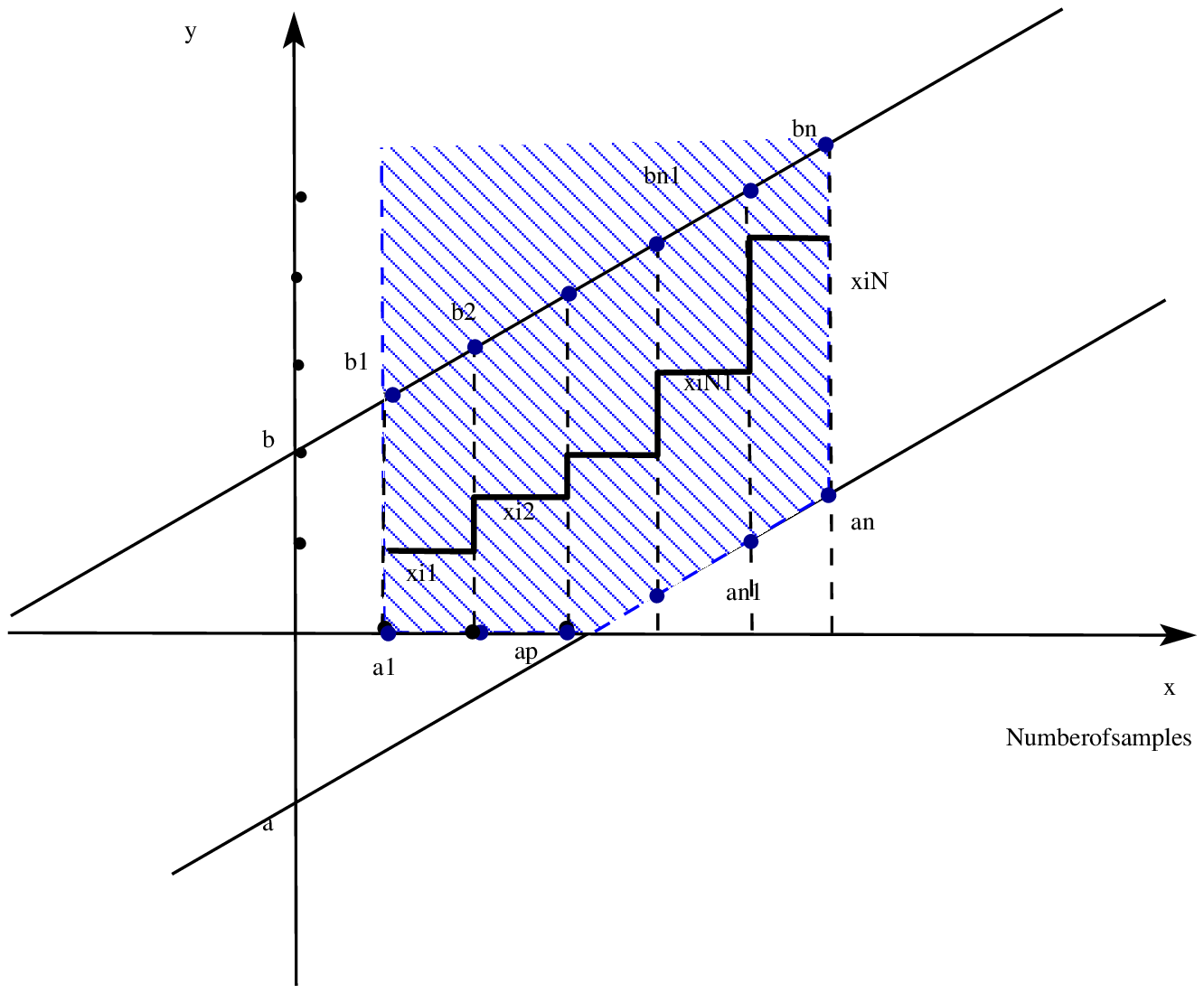,width=0.43\textwidth} \label{fig:rn1}}
\subfigure[Possible Paths from $\Xi_n^{(N)}$.]{
\psfrag{a1}{\footnotesize{$a_1$}}\psfrag{a2}{\footnotesize{${a_2}$}}\psfrag{an1}{\footnotesize{$a_{N-1}$}}
\psfrag{an}{\footnotesize{$a_N$}}
\psfrag{bsn1}{\footnotesize{$b_{n+1}$}}
\psfrag{b1}{\footnotesize{$b_1$}}\psfrag{ap}{\footnotesize{$a_P$}}
\psfrag{b2}{\footnotesize{$b_2$}}\psfrag{bn1}{\footnotesize{$b_{N-1}$}}\psfrag{bn}{\footnotesize{$b_{N}$}}
\psfrag{a}{\footnotesize{$a$}} \psfrag{b}{\footnotesize{$b$}}
\psfrag{x}{\footnotesize{$N$}} \psfrag{y}{\footnotesize{$\xi_N$}}
\psfrag{Numberofsamples}{\footnotesize{$\tt{Sample~Index}$}}
\psfrag{xi1}{\footnotesize{$\xi_1$}}\psfrag{xi2}{\footnotesize{$\xi_2$}}
\psfrag{xiN1}{\footnotesize{$\xi_{N-1}$}}\psfrag{xiN}{\footnotesize{$\xi_N$}}
\epsfig{file=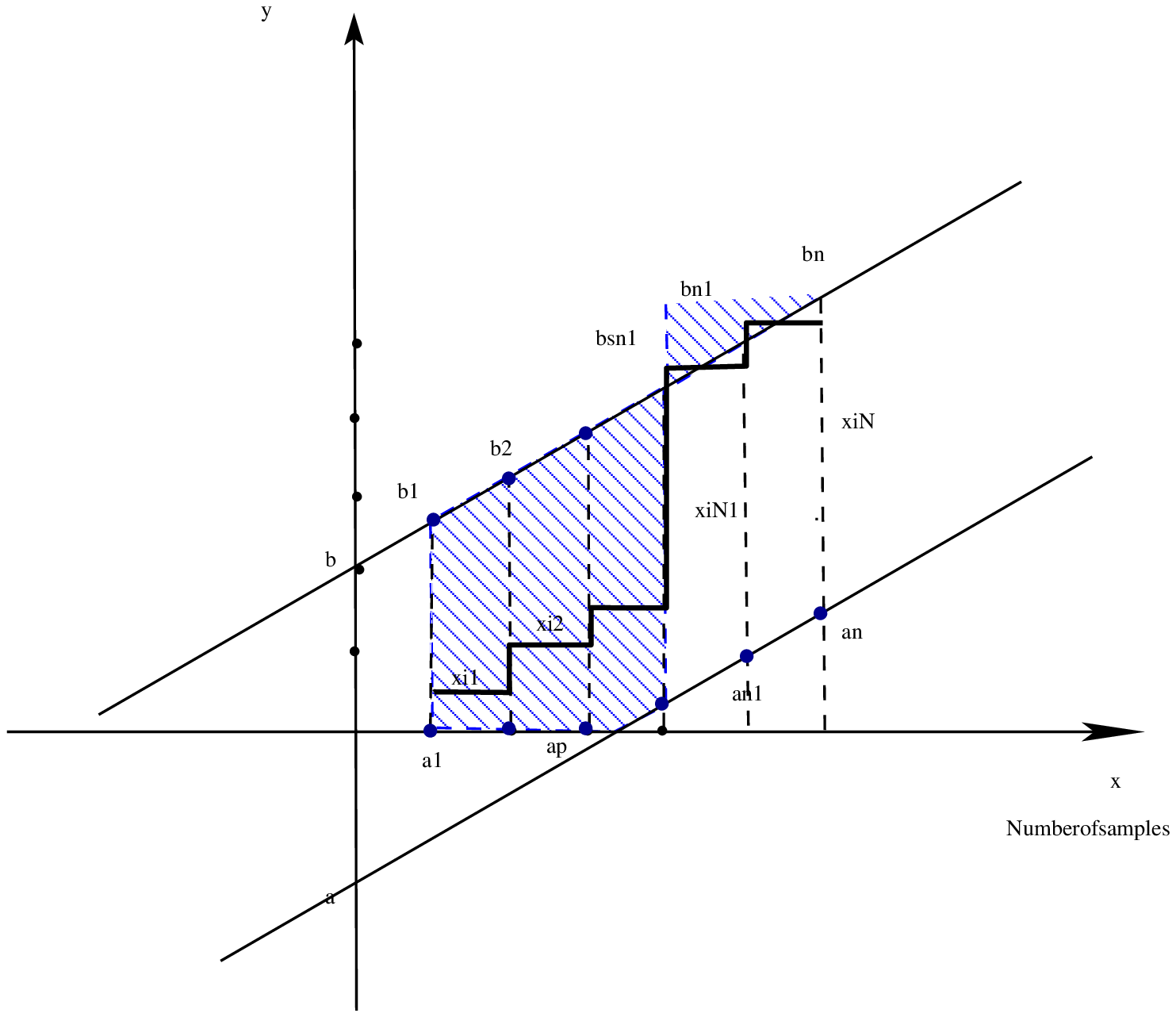,width=0.43\textwidth} \label{fig:rn2}}
\subfigure[Possible Paths from $\Omega^{(N)}~(N>Q)$.]{
\psfrag{a1}{\footnotesize{$a_1$}}\psfrag{a2}{\footnotesize{${a_2}$}}\psfrag{an1}{\footnotesize{$a_{N-1}$}}
\psfrag{an}{\footnotesize{$a_N$}} \psfrag{aQ}{\footnotesize{$a_Q$}}
\psfrag{aQ1}{\footnotesize{$a_{Q+1}$}}
\psfrag{b1}{\footnotesize{$b_1$}}\psfrag{ap}{\footnotesize{$a_P$}}
\psfrag{b2}{\footnotesize{$b_2$}}\psfrag{bn1}{\footnotesize{$b_{N-1}$}}\psfrag{bn}{\footnotesize{$b_{N}$}}
\psfrag{a}{\footnotesize{$a$}} \psfrag{b}{\footnotesize{$b$}}
\psfrag{x}{\footnotesize{$N$}}
\psfrag{Numberofsamples}{\footnotesize{$\tt{Sample~Index}$}}
\psfrag{y}{\footnotesize{$\Lambda_N$}}
\psfrag{xi1}{\footnotesize{$\xi_1$}}\psfrag{xi2}{\footnotesize{$\xi_2$}}
\psfrag{xiN1}{\footnotesize{$\xi_{N-1}$}}\psfrag{xiN}{\footnotesize{$\xi_N$}}
\epsfig{file=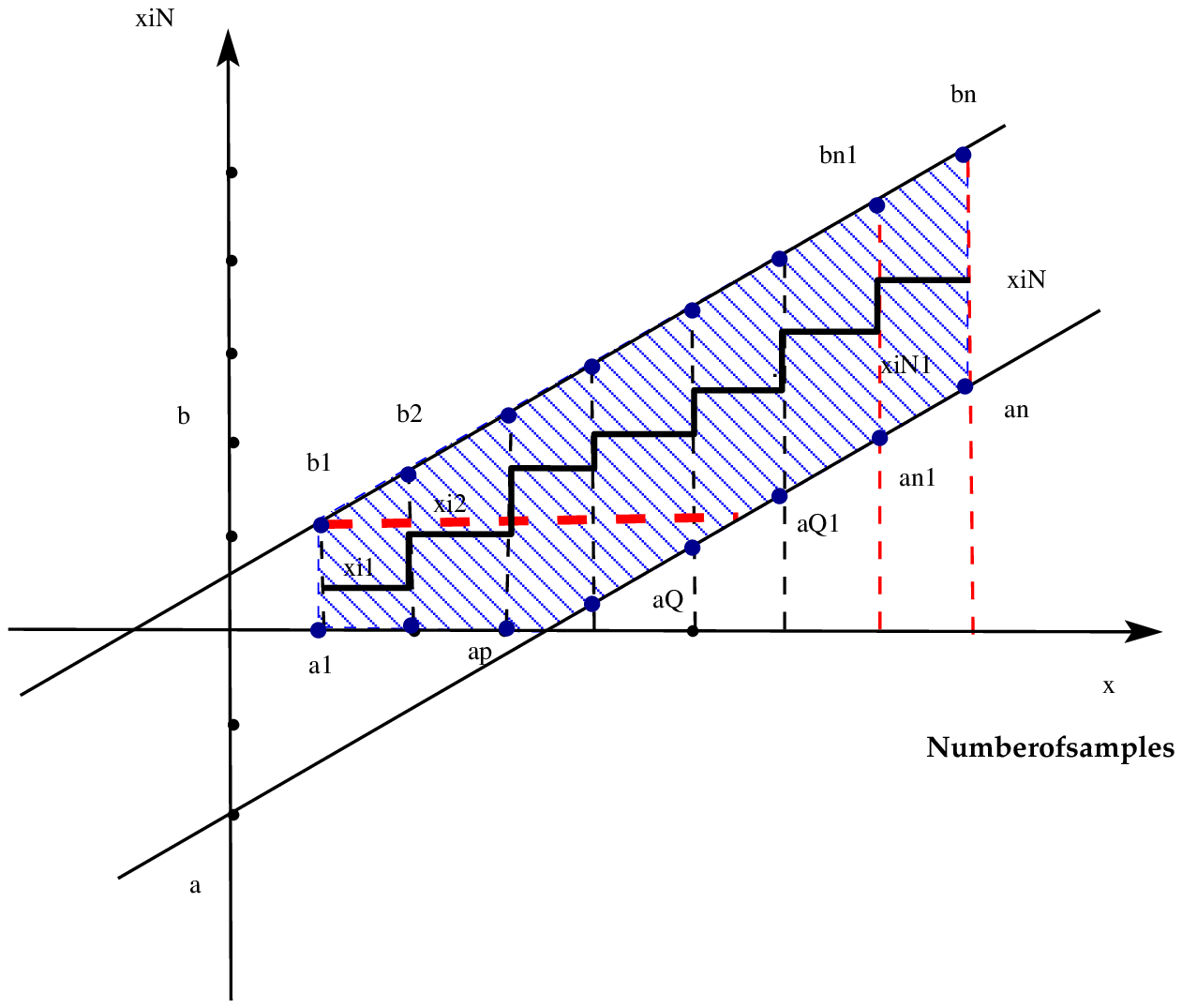,width=0.43\textwidth} \label{fig:rn3}}
\caption{An illustration for Proof of Lemma \ref{Lemma:lemma2}.}
\end{figure}

Recall that $\xi_N$ is a sum of $v_i$, i.e., $\xi_N=\sum_{i=1}^N
v_i$ with $v_i=2|r_i|^2/\sigma_w^2$. Hence,
$(\xi_1,\xi_2,\ldots,\xi_N)$ is a non-decreasing sequence, i.e.,
$\xi_0\le \xi_1 \le \xi_2 \le \cdots \le \xi_N$. Fig. \ref{fig:rn}
plots the parameter $\xi_N$ versus the sample index $N$. It is clear
from its definition that the region $\Omega^{(N)}$ contains all
possible sequences $(\xi_1,\xi_2,\ldots,\xi_N)$ (simply called paths
hereafter) satisfying $\xi_0\le \xi_1 \le \xi_2 \le \ldots \le
\xi_N$ and $0\le a_i < \xi_i < b_i$. Hence, the $i$th component of
each path $(\xi_1,\xi_2,\ldots,\xi_N)$ in $\Omega^{(N)}$ is
lower-bounded by the maximum of $\xi_{i-1}$ and $a_i$, and is
upper-bounded by the minimum of $\xi_{i+1}$ and $b_i$. The direct
computation of $I^{(N)}$ is highly complex
\cite{IEEEexample:article_exacttruncatedsequential} due to numerous
possibilities for lower- and upper-limits in the integral $I^{(N)}$.
Considering the fact that the integral $I^{(N)}$ can be readily
computed if either the lower- or upper-limit is a constant, we
express $\Omega^{(N)}$ into an equivalent set, over which the
integration can be readily computed in a recursive fashion, thereby
obviating the need to exhaustively enumerate these possibilities.

Let $\phi_i,~i\in \aleph_1^N$, denote a sequence of real numbers
with $0\le \phi_1 \le \phi_2 \le \ldots \le \phi_N$. Let us first
define the following set,
\begin{align}\label{eq:pinbalpha}
\Pi^{(N-n)}_{\bphi_n^N}=\{&(\xi_{n+1},\xi_{n+2},\ldots,\xi_N):\notag
\\ &0\le \xi_{n+1} \le \xi_{n+2} \ldots \le \xi_N; \phi_{i} < \xi_{i} \le
\xi_{i+1},i\in \aleph_{n+1}^{N-1};\phi_N < \xi_N < b_N \}
\end{align}
where $\bphi_n^N:=[\phi_{n+1},\ldots,\phi_N]$ is an
$(N-n)$-dimensional real vector with the $i$th entry of the vector
$\bphi_{n}^{N}$ being the lower bound of $\xi_{n+i}$ for $i\in
\aleph_1^{N-n}$. We next define the following non-overlapping
subsets of $\Pi_{\sbav_{0}^{N}}^{(N)}$,
\begin{align}\label{eq:xiNn}
\Xi^{(N)}_{n}:=\{&(\xi_1,\ldots,\xi_N):(\xi_1,\ldots,\xi_N)\in
\Pi_{\sbav_0^N}^{(N)}, a_i < \xi_i < b_i,i\in \aleph_1^{n};b_{n+1}
\le \xi_{n+1} \le \xi_{n+2}\}
\end{align}
where $n\in \aleph_0^{N-2}$.

As can be seen from Figs. \ref{fig:rn1} and \ref{fig:rn2},
$\Pi^{(N)}_{\sbav_0^{N}}$ contains all possible paths,
$(\xi_1,\xi_2,\ldots,\xi_N)$, which are lower-bounded by
$(a_1,a_2,\ldots,a_N)$ and upper-bounded by
$(\xi_2,\xi_3,\ldots,\xi_N,b_N)$, whereas $\Xi^{(N)}_{n}$ for $n\in
\aleph_0^{N-2}$ contains all possible paths
$(\xi_1,\xi_2,\ldots,\xi_N)$ having the property that the first $n$
variables lie in the set $\Omega^{(n)}$, i.e.,
$(\xi_1,\ldots,\xi_n)\in \Omega^{(n)}$, and the $(n+1)$th variable
$\xi_{n+1}$ excesses the upper slant line, i.e., $\xi_{n+1} \ge
b_{n+1}$. Again, it is clear from its definition that the set
$\Omega^{(N)}$ is equal to the difference between the set
$\Pi^{(N)}_{\sbav_{0}^N}$ and the union of $\Xi^{(N)}_{n}$ for $n\in
\aleph_0^{N-2}$, i.e., $\Omega^{(N)}=\Pi^{(N)}_{\sbav_0^N}\cap
\big(\cup_{n=0}^{N-2}\Xi^{(N)}_n\big)^c$. Thus, it follows from
\eqref{eq:integralI} that
\begin{equation}\label{eq:indifference}
I^{(N)}=\idotsint\limits_{\Pi^{(N)}_{\sbav_0^N}}d\bxi_N-
\sum_{n=0}^{N-2}\idotsint\limits_{\Xi^{(N)}_{n}}d\bxi_N.
\end{equation}

We now evaluate two terms on the right-hand side (RHS) of
\eqref{eq:indifference}. It is clear from \eqref{eq:giu} and
\eqref{eq:pinbalpha} that the first term on the RHS of
\eqref{eq:indifference} is nothing but $f_{\sbav_0^{N}}^{(N)}(b_N)$.
We next take a close look at the second term. The evaluation of the
second term is categorized into the following two cases:
\begin{itemize}
\item Case 1: $N\le Q$: In this case, we have $b_{n+1}> b_1\ge
a_n$ for any $n\in \{1,\ldots,N\}$. It implies from \eqref{eq:xiNn}
that we can express $\Xi^{(N)}_n$ as $\Xi^{(N)}_n=\Omega^{(n)}\times
[b_{n+1},\xi_{n+2}] \times \cdots \times [b_{n+1},\xi_N]\times
[b_{n+1},b_N]$ for $n=0,\ldots,N-2$ where $\Omega^{(0)}:=\emptyset$.
According to \eqref{eq:pinbalpha}, we have
\begin{equation}\label{eq:xiNndecompose}
\Xi^{(N)}_{n}=\Omega^{(n)}\times
\Pi^{(N-n)}_{b_{n+1}\bone_{N-n}},~n\in \aleph_0^{N-2}.
\end{equation}
Since $\xi_{n+1}> b_{n+1} > b_n> \xi_n$, the integral over
$\Omega^{(n)}$ and that over $\Pi^{(N-n)}_{b_{n+1}\bone_{N-n}}$ are
separable. Hence, relying on \eqref{eq:Ialphabetacase},
\eqref{eq:xiNn} and \eqref{eq:xiNndecompose}, we have
\begin{align}\label{eq:integralEncase1}
\idotsint\limits_{\Xi^{(N)}_n}d\bxi_N= \idotsint
\limits_{\Pi^{(N-n)}_{b_{n+1}\bone_{N-n}}}d\xi_{n+1}\cdots d\xi_N
\times \idotsint\limits_{\Omega^{(n)}}d\xi_1\cdots
d\xi_{n}=f^{(N-n)}_{b_{n+1}\bone_{N-n}}(b_N)I^{(n)}.
\end{align}

\item Case 2: $N\ge Q+1$:  The proof in this case follows the same
line of argument as that in the previous case. The key difference is
that because $N \ge Q+1$, some $a_n$ may be larger than $b_{n+1}$,
as depicted in Fig. \ref{fig:rn3}. To be specific, from the
definition of $Q$, we have
\[
a_Q \le b_1 < a_{Q+1},\ldots, a_{Q+n} \le b_{n+1} <
a_{Q+n+1},\ldots, a_{N} \le b_{N-Q+1} < a_{N+1}.
\]

1) For $n \in \aleph_0^{N-Q-1}$, we have
\[
\Xi^{(N)}_n=\Omega^{(n)} \times
\underset{Q}{\underbrace{[b_{n+1},\xi_{n+2}]\times \cdots \times
[b_{n+1},\xi_{n+Q+1}]}} \times \underset{N-Q-n}{\underbrace{
[a_{Q+n+1},\xi_{Q+n+2}] \times \cdots \times [a_{N-1},\xi_N]\times
[a_N,b_N]}}.
\]
Equivalently, $\Xi_n^{(N)}=\Omega^{(n)}\times
\Pi^{(N-n)}_{\bpsi_{n,a_N}^N}$ with
$\bpsi_{n,a_N}^N=[\underset{Q}{\underbrace{b_{n+1},\ldots,b_{n+1}}},
\underset{N-Q-n}{\underbrace{a_{Q+n+1},\ldots,a_N}}]$.

2) For~$n \in \aleph_{N-Q}^{N-2}$, we have $a_N \le b_{n+1}$. Due to
$a_i\le a_N$ for $i\in \aleph_1^{N-1}$, we have $a_i \le b_{n+1}$
for all $i\in \aleph_1^{N-1}$. Since any $(\xi_{n+1},\ldots,\xi_N)
\in \Pi^{(N-n)}_{\bpsi_{n,a_N}^N}$ belongs to a Cartesian product of
$(N-n)$ intervals (a hyper-rectangle) $[b_{n+1},\xi_{n+2}]\times
\ldots \times [b_{n+1},\xi_N]\times [b_ {n+1},b_N]$ having the same
lower limit $b_{n+1}$, this case is the same as Case 1.
Equivalently, $\Xi_n^{(N)}=\Omega^{(n)}\times
\Pi^{(N-n)}_{\bpsi_{n,a_N}^N}$ with
$\bpsi_{n,a_N}^N=b_{n+1}\bone_{N-n}$. Summarizing the preceding
results for Case 2, we have
\begin{equation}\label{eq:integralEncase2}
\idotsint\limits_{\Xi^{(N)}_n}d\bxi_N=\idotsint
\limits_{\Pi^{(N-n)}_{\bpsi_{n,a_N}^N}}d\xi_{n+1}\cdots d\xi_N
\times \idotsint\limits_{\Omega^{(n)}}d\xi_1\cdots
d\xi_{n}=f^{(N-n)}_{\bpsi_{n,a_N}^N}(b_N)I^{(n)}
\end{equation}
where
$\bpsi_{n,a_N}^N=[\underset{Q}{\underbrace{b_{n+1},\ldots,b_{n+1}}},\underset{N-Q-n}{\underbrace{a_{Q+n+1},\ldots,a_N}}]$
for $n \in \aleph_0^{N-Q-1}$ and
$\bpsi_{n,a_N}^N=b_{n+1}\bone_{N-n}$ for $n \in \aleph_{N-Q}^{N-2}$.
\end{itemize}

The proof follows immediately from \eqref{eq:indifference},
\eqref{eq:integralEncase1}, and \eqref{eq:integralEncase2} and the
fact that the first term on the RHS of \eqref{eq:indifference} is
equal to $f_{\sbav_0^{N}}^{(N)}(b_N)$. \hfill $\blacksquare$

\section{Proof of Lemma \ref{Lemma:lemma3}}\label{proof:lemma3}
Though the idea of the proof can be extended to a general case of
$a_{N-1} \le c$ and $a_N\le d$, we will consider the following two
cases: Case 1: $c:=\bar{\gamma}_N \le b_N$ and $d:=\infty$, and Case
2: $c:=a_{N}$ and $d:=b_{N}$, which correspond to
\eqref{eq:jnexpression1} and \eqref{eq:jnexpression3} respectively.
Define the following sets
\begin{align}
\Theta^{(N)}_{c,d}:=\{&(\xi_{1},\ldots,\xi_N): \xi_0\le \xi_1 \le
\ldots
\le \xi_N; a_i<\xi_i\le \xi_{i+1},i\in \aleph_1^{N-1},c<\xi_N <d \}, \\
\Phi_{c,d}^{(N,n)}:=\{&(\xi_1,\ldots,\xi_N): (\xi_1,\ldots,
\xi_{N})\in \Theta^{(N)}_{c,d};a_i < \xi_i < b_i,i \in
\aleph_1^n;b_{n+1} < \xi_{n+1} \le \xi_{n+2}\},
\end{align}
where  $\Phi_{c,d}^{(N,n)}$ are non-overlapping subsets of
$\Theta^{(N)}_{c,d}$ for $n\in \aleph_0^{N-2}$. The integral over
$\Theta^{(N)}_{c,d}$ can be readily computed as
\begin{align}\label{eq:integralthetan}
\idotsint\limits_{\Theta^{(N)}_{c,d}}e^{-\theta \xi_N}
d\bxi_N&=\int_{c}^{d} e^{-\theta \xi_N}
d\xi_N\int^{\xi_N}_{a_{N-1}}d\xi_{N-1}\cdots
\int^{\xi_2}_{a_1}d\xi_1 \notag \notag \\ &= \int_{c}^{d} e^{-\theta
\xi_N} f^{(N-1)}_{\sbav^{N-1}_0}(\xi_N)d\xi_N\notag \\ &
=\sum_{i=1}^N{\theta^{-i}}\big[f^{(N-i)}_{\sbav_0^{N-i}}(c)e^{-\theta
c}-f^{(N-i)}_{\sbav_0^{N-i}}(d)e^{-\theta d}\big]
\end{align}
where the equality in \eqref{eq:integralthetan} is obtained by using
integration by parts repeatedly and the differential property
$df^{(k)}_{\bchi_k}(\xi)/d\xi=f^{(k-1)}_{\bchi_{k-1}}(\xi)$.

\begin{itemize}
\item Case 1: $a_N\le \bar{\gamma}_N \le b_N$ and $ d=\infty$.
Similarly to the argument used in Lemma \ref{Lemma:lemma2}, we have
$\Upsilon^{(N)}_{\bar{\gamma}_N,\infty}=\Theta^{(N)}_{\bar{\gamma}_N,\infty}\bigcap
\big(\cup_{n=0}^{N-2}\Phi_{\bar{\gamma}_N,\infty}^{(N,n)}\big)^c$
and thus we have
\begin{equation}\label{eq:jcdn}
J_{\bar{\gamma}_N,\infty}^{(N)}(\theta)=\idotsint\limits_{\Theta^{(N)}_{\bar{\gamma}_N,\infty}}
e^{-\theta \xi_N}
d\bxi_N-\sum_{n=0}^{N-2}\idotsint\limits_{\Phi^{(N,n)}_{\bar{\gamma}_N,\infty}}
e^{-\theta \xi_N} d\bxi_N.
\end{equation}
Substituting $c=\bar{\gamma}_N$ and $d=\infty$ in
\eqref{eq:integralthetan} and using the fact that $e^{-\theta d}$ is
zero for $\theta>0$, we have
\[
\idotsint\limits_{\Theta^{(N)}_{\bar\gamma_N,\infty}}e^{-\theta
\xi_N}
d\bxi_N=\sum_{i=1}^N{\theta^{-i}}f^{(N-i)}_{\sbav^{N-i}_0}(\bar{\gamma}_N)e^{-\theta
\bar{\gamma}_N}.
\]

1) For $\bar{\gamma}_N \le b_1$, we have $b_1\ge a_{N}$ since
$\bar\gamma_N \ge a_N$. Since $\xi_n\le b_n < b_{n+1}$, we have
$\Phi_{\bar{\gamma}_N,\infty}^{(N,n)}=\Omega^{(n)}\times
\Pi^{(N-n)}_{b_{n+1}\bone_{N-n}}$ and the integrations over
$\Omega^{(n)}$ and $\Pi^{(N-n)}_{b_{n+1}\bone_{N-n}}$ are separable.
Thus, applying integration by parts and the fact
$f^{(k)}_{\bchi_k}(\chi_k)=0$, we obtain
\begin{align}\label{eq:rNlessb1}
\idotsint\limits_{\Phi^{(N,n)}_{\bar{\gamma}_N,\infty}} e^{-\theta
\xi_N} d\bxi_N&=\idotsint \limits_{\Pi^{(N-n)}_{b_{n+1}\bone_{N-n}}}
e^{-\theta \xi_N} d\xi_{n+1}\cdots d\xi_N \times
\idotsint\limits_{\Omega^{(n)}}d\xi_1\cdots d\xi_{n} \notag \\
&=I^{(n)}\int_{b_{n+1}}^\infty
e^{-\theta\xi_N}f_{b_{n+1}\bone_{N-n-1}}^{(N-n-1)}(\xi_N)d\xi_N
\notag \\
&=I^{(n)}\theta^{n-N}e^{-\theta b_{n+1}}.
\end{align}

2) For $\bar{\gamma}_N > b_1$, we have $s\ge 1$ since $b_s<
\bar{\gamma}_N \le b_{s+1}$. Similarly to the argument used in Case
2 of the proof of Lemma \ref{Lemma:lemma2}, we have
$\Phi_{n}^{(N)}=\Omega^{(n)}\times
\Pi^{(N-n)}_{{\bpsi}_{n,\bar{\gamma}_N}^N}$,
\begin{eqnarray}\label{eq:rNgreaterb1}
\idotsint\limits_{\Phi^{(N,n)}_{\bar{\gamma}_N,\infty}} e^{-\theta
\xi_N} d\bxi_N=\begin{cases} I^{(n)}\sum_{i=1}^{N-n}
\theta^{-i}f^{(N-n-i)}_{{\bpsi}_{n,\bar{\gamma}_N}^{N-i}}(\bar{\gamma}_N)e^{-\theta
\bar{\gamma}_N},&n\in \aleph_0^{s-1} \\ I^{(n)}\sum_{i=1}^{N-n}
\theta^{-i}f^{(N-n-i)}_{{\bpsi}_{n,\bar{\gamma}_N}^{N-i}}(b_{n+1})e^{-\theta
b_{n+1}} ,&n\in \aleph_s^{N-2}
\end{cases}.
\end{eqnarray}
This concludes the proof for Case 1.

\item Case 2: $c=a_{N}$ and $b=b_N$. Similarly to the method used in Case 1, we have
$\Upsilon^{(N)}_{a_N,b_N}=\Theta^{(N)}_{a_N,b_N}\bigcap
\big(\cup_{n=0}^{N-2}\Phi_{a_N,b_N}^{(N,n)}\big)^c$ and thus we
can express $J_{a_N,b_N}^{(N)}(\theta)$ as
\begin{align}\label{eq:jncase22}
J_{a_{N},b_N}^{(N)}(\theta)=\idotsint\limits_{\Theta^{(N)}_{a_{N},b_N}}
e^{-\theta \xi_N} d\bxi_N-{\mathbb{I}}_{\{N\ge
2\}}\sum_{n=0}^{N-2}\idotsint\limits_{\Phi^{(N,n)}_{a_{N},b_N}}
e^{-\theta \xi_N} d\bxi_N.
\end{align}
The rest of the proof is analogous to that in Case 1. The key
difference is that in Case 2, the term $e^{-\theta d}$ is no longer
zero. From \eqref{eq:integralthetan}, the first term on the RHS of
\eqref{eq:jncase22} can be readily obtained as
\begin{align}\label{eq:thetancase22}
\idotsint\limits_{\Theta^{(N)}_{a_{N},b_N}}e^{\theta \xi_N} d\bxi_N
=\sum_{i=1}^N\theta^{-i}\big[f^{(N-i)}_{\sbav_0^{N-i}}(a_{N})e^{\theta
a_{N}}-f^{(N-i)}_{\sbav_0^{N-i}}(b_N)e^{-\theta
 b_{N}}\big].
\end{align}

Since $b_{N-Q} < a_N \le b_{N-Q+1}$, we have $s=N-Q$ in this case.
Similarly to the method used to derive \eqref{eq:rNlessb1} and
\eqref{eq:rNgreaterb1}, we have
\begin{align}\label{eq:phincase22}
&\idotsint\limits_{\Phi^{(N,n)}_{a_{N},b_N}} e^{-\theta \xi_N}
d\bxi_N\notag \\&=\begin{cases} \displaystyle I^{(n)}
\Big[\theta^{n-N}e^{-\theta b_{n+1}}-\sum_{i=1}^{N-n} \theta^{-i}
f^{(N-n-i)}_{b_{n+1}\bone_{N-n-i}}(b_N)e^{-\theta d}\Big],a_N \le b_1,n\in \aleph_0^{N-2} \\
\displaystyle I^{(n)}\sum_{i=1}^{N-n} \theta^{-i}\Big[
f^{(N-n-i)}_{{\bpsi}_{n,a_N}^{N-i}}(a_N)e^{-\theta
a_N}-f^{(N-n-i)}_{{\bpsi}_{n,a_N}^{N-i}}(b_N)e^{-\theta b_N}\Big]
,a_N> b_1,n\in \aleph_0^{N-Q-1} \\
\displaystyle I^{(n)}\Big[\theta^{n-N}e^{-\theta
b_{n+1}}-\sum_{i=1}^{N-n}
\theta^{-i}f^{(N-n-i)}_{b_{n+1}\bone_{N-n-i}}(b_N)e^{-\theta
b_N}\Big],a_N> b_1,n\in \aleph_{N-Q}^{N-2}.
\end{cases}
\end{align}

The proof for Case 2 follows clearly from \eqref{eq:jncase22},
\eqref{eq:thetancase22}, and \eqref{eq:phincase22}. \hfill
$\blacksquare$
\end{itemize}

\section{Proof of Property
\ref{property2}}\label{proof:property2}

Note that $F_N=\mathcal{A}_N^{l_N}\cap \mathcal{B}_N^{l_N}$ and
$\tilde F_N=\tilde {\mathcal{A}}^{l_N}_N \cap
\tilde{\mathcal{B}}^{l_N}_N$. Applying the inclusion-exclusion
identity \cite[p. 80]{FristedtGray_probability_book}, we have
$P_{H_1}(F_N)=P_{H_1}(\mathcal{A}_N^{l_N})+P_{H_1}(\mathcal{B}_N^{l_N})-P_{H_1}(\mathcal{A}_N^{l_N}\cup
\mathcal{B}_N^{l_N})$ and $P_{H_1}(\tilde F_N)=P_{H_1}(\tilde
{\mathcal{A}}_N^{l_N})+P_{H_1}(\tilde{\mathcal{B}}_N^{l_N})-P_{H_1}(\tilde
{\mathcal{A}}_N^{l_N}\cup \tilde{\mathcal{B}}_N^{l_N})$. Thus, by
using the triangle inequality, we have
\begin{align}\label{eq:ph1ph1tilde}
|P_{H_1}(F_N)-P_{H_1}(\tilde F_N)|  {\le} &
|P_{H_1}({\mathcal{A}}_N^{l_N})-P_{H_1}(\tilde
{\mathcal{A}}_N^{L_N})|+|P_{H_1}(\mathcal{B}_N^{L_N})-P_{H_1}(\tilde{\mathcal{B}}_N^{L_N})|\notag
\\ &+|P_{H_1}(\mathcal{A}_N^{L_N}\cup
\mathcal{B}_N^{L_N})-P_{H_1}(\tilde{\mathcal{A}}_N^{L_N} \cup \tilde
{\mathcal{B}}_N^{L_N})|.
\end{align}

Since $1-\epsilon/(3M) \le P_{H_1}(\mathcal{A}_N^{L_N}) \le 1$ and
$1-\epsilon/(3M) \le P_{H_1}(\tilde{\mathcal{A}}_N^{L_N}) \le 1$, we
have $1-\epsilon/(3M) \le P_{H_1}(\mathcal{A}_N^{L_N}) \le
P_{H_1}(\mathcal{A}_N^{L_N}\cup \mathcal{B}_N^{L_N}) \le 1$ and
$1-\epsilon/(3M) \le P_{H_1}(\tilde{\mathcal{A}}_N^{L_N}) \le
P_{H_1}(\tilde{\mathcal{A}}_N^{L_N} \cup \tilde
{\mathcal{B}}_N^{L_N}) \le 1$. It can be readily inferred from the
above inequalities that
$|P_{H_1}(\mathcal{A}_N^{L_N})-P_{H_1}(\tilde{\mathcal{A}}_N^{L_N})|
\le \epsilon/(3M)$ and $|P_{H_1}(\mathcal{A}_N^{L_N}\cup
\mathcal{B}_N^{L_N})-P_{H_1}(\tilde{\mathcal{A}}_N^{L_N}\cup
\tilde{\mathcal{B}}_N^{L_N})| \le \epsilon/(3M)$. This, along with
the inequality \eqref{eq:ph1ph1tilde} and the assumption
$|P_{H_1}(\mathcal{B}_N^{L_N})-P_{H_1}(\tilde{\mathcal{B}}_N^{L_N})|\le
\epsilon/(3M)$, implies that $|P_{H_1}(F_N)-P_{H_1}(\tilde F_N)| <
\epsilon/M$. Hence, we have
\[
|\beta_{\text{ssct}}-\tilde{\beta}_{\text{ssct}}| \le \sum_{i=1}^M
|P_{H_1}(F_N)-P_{H_1}(\tilde F_N)| \le \epsilon.
\].
\hfill $\blacksquare$

\begin{table}
\caption{SCCT Versus Energy Detection} \centerline{
\begin{tabular}{c|c|c|c|c}
\hline  ${\tt{SNR}}_{m}~\text{(dB)} $ & $0$  & $-5$  & $-10$  &
$-15$
\\
\hline
$\bar\gamma$ & $-8.5$ & $-5.69$ & $-4$ & $-1.897$ \\
$\bar b$ & $27$ & $35.32$ & $69.30$ & $158.47$\\
$\bar{\Delta}$ & $3$ & $2.316$ & $2.100$ & $2.032$ \\
\hline
$\alpha_{\text{ssct}}~(\text{Monte Carlo})$ & $0.011$ & $0.055$& $0.103$ & 0.153\\
$\alpha_{\text{ssct}}~(\text{Numerical})$  & $0.011$ & $0.055$ & $0.103$ & 0.153\\
$\alpha_{\text{ed}}~(\text{Energy Detect.})$ & $0.011$ & $0.055$ & $0.101$ & 0.150\\
\hline
$\beta_{\text{ssct}}~(\text{Monte Carlo})$  & $0.008$ & $0.046$ & $0.099$ & 0.154\\
$\beta_{\text{ssct}}~(\text{Numerical})$  & $0.008$ & $0.047$ & $0.100$ & 0.156\\
$\beta_{\text{ed}}~(\text{Energy Detect.})$ & $0.008$ & $0.046$ & $0.096$ & 0.149 \\
\hline
$\text{ASN}~(\text{Monte Carlo})$ & $26$ & $95$ & $509$ & $3154$\\
$\text{ASN}~(\text{Numerical})$ & $26$ & $96$ & $515$ & $3185$\\
$M$ ($\text{Energy Detect.}$) & $40$ & $140$ & $730$ & $4450$  \\
\hline $\text{Efficiency}~\eta$ & $35\%$ & $32\%$ &
 $30\%$ & $29\%$
\\
\hline
\end{tabular}}
\label{table1:simulation}
\end{table}

\begin{table}
\caption{Detection Performance Without Knowing Modulation Types of
the Primary Signals} \centerline{
\begin{tabular}{c|c|c|c|c}
\hline  $\text{SNR}_{m}$ (dB) & $0$  & $-5$ & $-10$ & $-15$
\\
\hline
$\beta_{\text{ssct}}~(\text{QPSK,Monte Carlo})$ & $0.008$ & $0.046$ & $0.099$ & $0.154$\\
$\beta_{\text{ssct}}~(\text{QPSK,Numerical})$ & $0.008$ & $0.047$ & $0.100$ & $0.156$\\
$\beta_{\text{ed}}~(\text{QPSK,Energy Detect.})$ & $0.008$ & $0.046$ & $0.096$ & $0.149$\\
$\beta_{\text{ssct}}~(\text{64-QAM,Monte Carlo})$ & $0.012$ & $0.048$ & $0.099$ & $0.154$ \\
$\beta_{\text{ssct}}~(\text{64-QAM,Numerical})$ & $0.012$ & $0.050$ & $0.103$ & $0.157$\\
$\beta_{\text{ssct}}~(\text{64-QAM,Energy Detect.})$ & $0.012$ & $0.047$ & $0.096$ & $0.149$\\
\hline
$\text{ASN}~(\text{QPSK,Monte Carlo})$ & $26$ & $95$ & $509$ & $3154$   \\
$\text{ASN}~(\text{QPSK,Numerical})$ & $26$ & $96$ & $515$ & $3185$ \\
$\text{ASN}~(\text{64-QAM,Monte Carlo})$ & $26$ & $95$ & $509$ & $3154$ \\
$\text{ASN}~(\text{64-QAM,Numerical})$ & $26$ & $96$& $514$ & $3190$ \\
\hline
\end{tabular}}
\label{table2:simulation}
\end{table}

\begin{table}
\caption{Mismatch between $\text{SNR}_m$ and $\text{SNR}_o$
($\text{SNR}_m=-15$ ${\tt{dB}}$)} \centerline{
\begin{tabular}{c|c|c|c|c}
\hline  $\text{SNR}_{o}$ (dB) & $-12$  & $-13$ & $-14$ & $-15$\\
\hline
$\beta_{\text{ssct}}~(\text{Monte Carlo})$ & $0.0018$ & $0.0153$ & $0.0629$ & $0.154$ \\
$\beta_{\text{ssct}}~(\text{Numerical})$ & $0.0017$ & $0.0151$ & $0.0628$ & $0.156$ \\
$\beta_{\text{ed}}~(\text{Energy Detect.})$ & $0.0012$ & $0.0131$ & $0.0584$ & $0.149$ \\
\hline
$\text{ASN}~(\text{Monte Carlo})$ & $2425$ & $2686$ & $2948$ & $3154$    \\
$\text{ASN}~(\text{Numerical})$ & $2499$ & $2769$ & $3035$ & $3185$  \\
$M~\text{(Energy Detect.)}$ & $4450$ & $4450$ & $4450$ & $4450$\\
\hline
$\text{Efficiency}~(\eta)$ & $46\%$ & $40\%$ & $34\%$ & $29\%$\\
\hline
\end{tabular}}
\label{table3:simulation}
\end{table}

\begin{table}
\caption{Impacts of Truncation Size $M$ ($\text{SNR}=-5~{\tt{dB}}$,
Monte Carlo Simulation)} \centerline{
\begin{tabular}{c|c|c|c|c|c|c}
\hline  $M$ & $\bar a$ & $\bar b$  & $\bar \gamma$ & $\text{ASN}$ & $T_p$  & $\eta$\\
\hline
$M=140$ & $-35.32$ & $35.32$ & $-5.69$ & $95.4$ & $26.8\%$ & $32\%$\\
$M=160$ & $-28.95$ & $23.16$ & $-5.50$ & $76.7$ & $9.2\%$  & $45\%$\\
$M=180$ & $-27.33$ & $21.54$ & $-6.00$ & $73.1$ & $5.1\%$ & $48\%$\\
$M=200$ & $-26.40$ & $20.85$ & $-6.32$ & $71.2$ & $3.0\%$ & $49\%$ \\
$M=500$ & $-25.48$ & $19.69$ & $-6.32$ & $68.8$ & $0.005\%$ & $51\%$ \\
$M=1000$ & $-25.42$ & $19.63$ & $-6.32$ & $68.6$ & $0$ & $51\%$ \\
\hline $\text{SPRT}$ (non-truncated) & $-$ & $-$ & $-$ & $67.9$ &
$0$ & $52\%$
\\
\hline
\end{tabular}}
\label{table4:simulation}
\end{table}

\bibliographystyle{IEEEtran}
\bibliography{IEEEabrv,mybib}

\begin{thebibliography}{10}
\providecommand{\url}[1]{#1}
\csname url@rmstyle\endcsname
\providecommand{\newblock}{\relax}
\providecommand{\bibinfo}[2]{#2}
\providecommand\BIBentrySTDinterwordspacing{\spaceskip=0pt\relax}
\providecommand\BIBentryALTinterwordstretchfactor{4}
\providecommand\BIBentryALTinterwordspacing{\spaceskip=\fontdimen2\font plus
\BIBentryALTinterwordstretchfactor\fontdimen3\font minus
  \fontdimen4\font\relax}
\providecommand\BIBforeignlanguage[2]{{%
\expandafter\ifx\csname l@#1\endcsname\relax
\typeout{** WARNING: IEEEtran.bst: No hyphenation pattern has been}%
\typeout{** loaded for the language `#1'. Using the pattern for}%
\typeout{** the default language instead.}%
\else
\language=\csname l@#1\endcsname
\fi
#2}}

\bibitem{FCC2002:reportcognitive}
{{Federal Communications Commission}}, ``Spectrum policy task force,''
  \emph{Rep. ET Docket}, pp. 1--135, Nov. 2002.

\bibitem{Liang:tradeoff08}
Y.-C. Liang, Y.~Zeng, E.~Peh, and A.~Hoang, ``Sensing-throughput tradeoff for
  cognitive radio networks,'' \emph{{IEEE} Trans. Wireless Commun.}, vol.~7,
  no.~4, pp. 1326--1337, Apr. 2008.

\bibitem{Cui07:jssp}
Z.~Quan, S.~Cui, and A.~Sayed, ``Optimal linear cooperation for spectrum
  sensing in cognitive radio networks,'' \emph{IEEE J. Select. Topics in Signal
  Processing}, vol.~2, no.~1, pp. 28--40, Feb. 2008.

\bibitem{seunggg:sqofdm}
S.~J. Kim and G.~B. Giannakis, ``Rate-optimal and reduced-complexity sequential
  sensing algorithms for cognitive ofdm radios,'' in \emph{Proc. of 43rd Conf.
  on Info. Sciences and Systems}, Johns Hopkins Univ., Baltimore, MD, Mar. 18
  -- 20, 2009.

\bibitem{laifanpoor2008}
L.~Lai, Y.~Fan, and H.~Poor, ``Quickest detection in cognitive radio: A
  sequential change detection framework,'' in \emph{Proc. {IEEE} Global
  Telecommunications Conference (GLOBECOM)}, New Orleans, LA, Nov. 2008, pp.
  1--5.

\bibitem{conf_icc_chengaodaut}
H.~S. Chen, W.~Gao, and D.~G. Daut, ``Signature based spectrum sensing
  algorithms for {IEEE} 802.22 {WRAN},'' in \emph{Proc. {IEEE} International
  Conference on Communications (ICC)}, Beijing, China, June 2008, pp.
  6487--6492.

\bibitem{article_energydetectionoriginalpaper}
H.~Urkowitz, ``Energy detection of unknown deterministic signals,'' \emph{Proc.
  {IEEE}}, vol.~55, no.~4, pp. 523--531, Apr. 1967.

\bibitem{conf_crowncom_lunden}
J.~Lund$\acute{\text{e}}$n, V.~Koivunen, A.~Huttunen, and H.~V. Poor,
  ``Spectrum sensing in cognitive radios based on multiple cyclic
  frequencies,'' in \emph{Proc. {IEEE} Cognitive Radio Oriented Wireless
  Networks and Communications (CrownCom)}, Orlando, FL, Aug. 2007, pp. 37--43.

\bibitem{journal:sahai08j}
R.~Tandra and A.~Sahai, ``{SNR} walls for signal detection,'' \emph{IEEE
  Journal of Selected Topics in Signal Processing}, vol.~2, no.~1, pp. 4--17,
  Feb. 2008.

\bibitem{icecs07:Kundargiewfik}
N.~Kundargi and A.~Tewfik, ``Hierarchical sequential detection in the context
  of dynamic spectrum access for cognitive radios,'' in \emph{Proc. IEEE 14th
  Int. Conf. on Electronics, Circuits and Systems}, Marrakech, Morocco, Dec.
  11-14, 2007, pp. 514--517.

\bibitem{chen06:technicalreport}
B.~Chen, J.~Park, and K.~Bian, ``Robust distributed spectrum sensing in
  cognitive radio networks,'' \emph{Technical Report TR-ECE-06-07, Dept. of
  Electrical and Computer Engineering, Virginia Tech}, July 2006.

\bibitem{IEEEexample:article_sequentialdetection}
A.~Wald, ``Sequential tests of statistical hypothesis,'' \emph{Ann. Math.
  Stat.}, vol.~17, pp. 117--186, 1945.

\bibitem{WaldWolfowitz:optimumcharacter}
A.~Wald and J.~Wolfowitz, ``Optimum character of the sequential probability
  ratio test,'' \emph{Ann. Math. Stat.}, vol.~19, pp. 326--329, 1948.

\bibitem{pollockgolhar:technicalreport}
S.~M. Pollock and D.~Golhar, ``Efficient recursions for truncation of the
  {SPRT},'' \emph{Technical Report No. 85-24, Dept. of Industrial and
  Operations Engineering, University of Michigan}, Aug. 1985.

\bibitem{IEEEexample:article_exacttruncatedsequential}
R.~C. Woodall and B.~M. Kurkjian, ``Exact operating characteristic for
  truncated sequential life tests,'' \emph{Ann. Math. Stat.}, vol.~33, pp.
  1403--1412, 1962.

\bibitem{Aroian68:directmethod}
L.~A. Aroian, ``Sequential analysis, direct method,'' \emph{Technometrics},
  vol.~10, no.~1, pp. 125--132, Feb. 1968.

\bibitem{surveyjohnson:sequential}
N.~L. Johnson, ``Sequential analysis: A survey,'' \emph{Journal of the Royal
  Statistical Society. Series A (General)}, no.~3, pp. 372--411, 1961.

\bibitem{limth08}
T.~H. Lim, R.~Zhang, Y.-C. Liang, and H.~Zeng, ``{GLRT}-based spectrum sensing
  for cognitive radio,'' in \emph{Proc. {IEEE} Global Telecommunications
  Conference (GLOBECOM)}, New Orleans, LA, Nov. 2008, pp. 1--5.

\bibitem{FristedtGray_probability_book}
B.~Fristedt and L.~Gray, \emph{A Modern Approach to Probability Theory}.\hskip
  1em plus 0.5em minus 0.4em\relax Boston: Birkh$\ddot{\text{a}}$user, 1997.

\end{thebibliography}

\end{document}